\documentclass[a4paper,12pt]{article}

\usepackage{amsmath}
\usepackage{float}
\usepackage{gensymb}
\usepackage{eurosym}
\usepackage{comment}
\usepackage{fancyvrb}
\usepackage{graphicx}
\usepackage{hyperref}
\usepackage[ragged]{footmisc}
\usepackage{ragged2e}
\usepackage[section]{placeins}
\usepackage{appendix}
\usepackage{geometry}
\usepackage{amsmath}
\usepackage{underscore}
\usepackage[english]{babel}

\begin{document}

\title{Problems in AI research and how the SP System may solve them}

\author{J Gerard Wolff}

\author{J Gerard Wolff\protect\footnote{Dr Gerry Wolff, BA (Cantab) PhD (Wales) CEng MBCS MIEEE, ORCID ID 0000-0002-4624-8904. CognitionResearch.org, Menai Bridge, UK; \href{mailto:jgw@cognitionresearch.org}{jgw@cognitionresearch.org}; +44 (0) 1248 712962; +44 (0) 7746 290775; {\em Twitter:} @gerrywolff65; {\em URL:} \href{http://www.cognitionresearch.org}{www.cognitionresearch.org}.}}

\maketitle

\begin{abstract}

This paper describes problems in AI research and how the SP System may solve them. The SP System is described in an appendix and sources referenced there. Most of the problems considered in the paper are described by leading researchers in AI in interviews with science writer Martin Ford, and reported by him in his book {\em Architects of Intelligence}. The problems that are considered are: the need to bridge the divide between symbolic and non-symbolic kinds of knowledge and processing; the tendency of deep neural networks (DNNs) to make large and unexpected errors in recognition; the need to strengthen the representation and processing of natural languages; the challenges of unsupervised learning; the need for a coherent account of generalisation, under-generalisation, and over-generalisation; how to learn usable knowledge from a single exposure or experience; how to achieve transfer learning (incorporating old knowledge in new); how to increase the speed of learning in AI systems, and how to reduce the demands of AI learning for large volumes of data, and for large computational resources; the need for transparency in the representation and processing of knowledge; how to achieve human-like capabilities in probabilistic reasoning; the need to re-balance research towards top-down strategies, with the need to develop `broad AI' (or `artificial general intelligence'); how to minimise the risk of accidents with self-driving vehicles; the need for strong compositionality in the structure of knowledge; the challenges of commonsense reasoning and commonsense knowledge; establishing the key importance of information compression in AI research; establishing the importance of a biological perspective in AI research; establishing whether knowledge in the brain is represented in `distributed' or `localist' form; how to bypass the limited scope for adaptation in deep neural networks; and how to eliminate the problem of catastrophic forgetting. Tbe evidence for how the SP System may solve these problems in AI research, with evidence for the versatility of the SP System, suggests that it provides a firmer foundation for the development of human-like general AI or `AGI' than any alternative. In this research, no attempt has been made to study the important areas of motivation and emotion.

\end{abstract}


\section{Introduction}\label{introduction-section}

This paper is about significant problems in research into artificial intelligence (AI), with evidence to show how the {\em SP System} may solve them.

The SP System---which means the {\em SP Theory of Intelligence} and its realisation in the {\em SP Computer Model} (SPCM)---is introduced in Section \ref{intro-to-the-sp-system-section}, below, with a fairly full outline in Appendix \ref{outline-of-sp-system-appendix}, and pointers to where more detail may be found. Reasons for the adoption of the name `SP' are given in Appendix \ref{simplicity-power-appendix}.

Most of the problems in AI research that are the subject of this paper have been described by leading researchers in AI in interviews with science writer Martin Ford, and reported by him in his book {\em Architects of Intelligence} \cite{ford-2018}. Here, those problems are discussed in Sections \ref{symbolic-non-symbolic-section} to \ref{distributed-localist-representations-section}, inclusive. The remaining problems are discussed in Sections \ref{scope-for-adaptation-section} to \ref{catastrophic-forgetting-section}, inclusive. And Section \ref{motivations-emotions-section} says merely that the important topic of motivations and emotions are, so far, outside the scope of the SP research.

It must be stressed that {\em this paper is {\bf not} a review of Ford's book}. It serves here simply as {\em a valuable source of information about problems in AI research that have been identified by leading researchers in AI}.

It must also be stressed that the paper is {\em not} an ``exposition type of paper''. It provides solid evidence about how the SP System may solve those problems in AI research.

As described in Section \ref{devt-of-sp-system-section}, below, the SP Theory of Intelligence, was developed in conjunction with the creation and testing of a very large number of versions of the SPCM over a period of about 17 years. As noted below, {\em that lengthy process of development has been extremely important in weeding out bad ideas and blind alleys in the evolving SP Theory of Intelligence}.

Appendix \ref{mathematics-in-sp-system-appendix} describes the mathematics incorporated in the SPCM, and mathematics that has contributed to its development.

Each of the afore-mentioned problems in AI research is described in its own section (Sections \ref{symbolic-non-symbolic-section} to \ref{catastrophic-forgetting-section}), with a description of how the SP System may solve it.

Where it is appropriate, each section contains a subsection headed `demonstration(s)' which presents one or more relevant demonstrations from the SPCM.
Where relatively full demonstrations have been published already, there is an outline of what those demonstrations show, and a pointer or pointers to relevant sources.

{\em The evidence presented in this paper, showing how the SP System may solve several different problems in AI research, with evidence for the versatility of the SP System, suggests that it provides a firmer foundation for the development of human-like general AI (or `AGI') than any alternative}.

\subsection{Abbreviations}\label{abbreviations-section}

Abbreviations used in this paper are detailed below.

\begin{itemize}

    \item {\em Artificial General Intelligence}: `AGI'.

    \item {\em Artificial Intelligence}: `AI'.

    \item {\em Commonsense Reasoning and Commonsense Knowledge}: `CSRK'.

    \item {\em Deep Neural Network}: `DNN'.

    \item {\em Human Learning, Perception, and Cognition}: `HLPC'.

    \item {\em Information Compression}: `IC'.

    \item {\em Information Compression Via the Matching and Unification of Patterns}: `ICMUP'.

    \item {\em `Simplification and Integration of Observations and Concepts Across AI, Mainstream Computing, Mathematics, and HLPC}: This expression may be shortened to {\em Simplification and Integration Across a Broad Canvass}, or SIABC'.

    \item {\em SP Computer Model}: `SPCM'.

    \item {\em SP-multiple-alignment}: `SPMA'.

\end{itemize}

It is intended that `SP' should be treated as a name. Reasons for that name are given in Appendix \ref{simplicity-power-appendix}.

\section{Preamble}\label{preamble-section}

This section describes aspects of the SP programme of research which underpin later sections of the paper.

\subsection{Introduction to the SP System}\label{intro-to-the-sp-system-section}

As noted above, the {\em SP System} means the {\em SP Theory of Intelligence} and its realisation in the {\em SP Computer Model}. To be more precise, the SP Theory of Intelligence is essentially the theory as it is described in words and diagrams, and the SP Computer Model provides great precision, and the means of testing the theory and demonstrating what it can do.

Those two parts of the SP System are the products of an extended programme of research, seeking {\em Simplification and Integration of Observations and Concepts Across AI, Mainstream Computing, Mathematics, and Human Learning, Perception, and Cognition}. This is broadly in accord with the need for top-down strategies in AI research (Section \ref{top-down-strategy-section}), and with the goal of developing human-like general intelligence or `artificial general intelligence' (AGI) (Section \ref{top-down-strategy-section}).

For the sake of brevity, ``Human Learning, Perception, and Cognition'' will be referred to as `HLPC', and ``Simplification and Integration of Observations and Concepts Across AI, Mainstream Computing, Mathematics, and HLPC'' may be shortened to ``Simplification and Integration Across a Broad Canvass'' or `SIABC'.

In this paper, the SP System is described in outline in Appendix \ref{outline-of-sp-system-appendix}, with enough detail to ensure that the rest of the paper makes sense.

In view of the importance of mathematics in the analysis of problems in this area, Appendix \ref{mathematics-in-sp-system-appendix} describes the mathematics that is embedded in the SPCM, and mathematics which has contributed to its development.

Elsewhere, the SP System is described quite fully in \cite{sp-extended-overview}, and even more fully in the book {\em Unifying Computing and Cognition} \cite{wolff-2006}. Other publications in this programme of research are detailed, with download links, \linebreak
on \href{http://www.cognitionresearch.org/sp.htm}{www.cognitionresearch.org/sp.htm}, with links to a selection of key publications \linebreak
on \href{http://www.cognitionresearch.org/extras/key_publications.htm}{www.cognitionresearch.org/extras/key-publications.htm}.

\subsection{Development of the SP Theory of Intelligence with the SP Computer Model}\label{devt-of-sp-system-section}

Confidence in the SP System may be strengthened by knowing how the SP Theory of Intelligence and the SPCM have been developed.

As described above, the aim has been to discover a framework for SIABC. From early ideas about how this may be done (see, for example, \cite{wolff-1990}), the process of creating a conceptual framework and developing it was done via the development and testing of a very large number of versions of the SPCM. At all stages, SIABC has been the touchstone of success or failure. The book {\em Unifying Computing and Cognition} \cite{wolff-2006} provides a full account of the framework and what it can do.

That process of searching for a good framework for SIABC, and developing and testing the SPCM, took about 17 years! {\em That lengthy process of development has been extremely important in weeding out bad ideas and blind alleys in the evolving SP Theory of Intelligence}, and it is largely responsible for the intelligence-related strengths and potential of the SP System, summarised in Appendix \ref{sp-str-pot-appendix}. It is also largely responsible for the several potential benefits and applications of the SP System, summarised in Appendix \ref{benefits-and-applications-appendix}.

Thus {\em the SPCM was not hacked together in a day}. Its long period of research, development, and testing, provides a solid foundation for arguments and examples later in the paper, relating to the structure and workings of SP System.

\subsection{Information compression and the SP-multiple-alignment construct}\label{ic-spma-section}

On the strength of substantial evidence for the importance of information compression (IC) in human learning, perception, and cognition (HLPC) \cite{sp-compression}, IC is fundamental in the workings of the SP System (Appendices \ref{high-level-view-of-sp-appendix} and \ref{ic-via-icmup-appendix}).

More specifically, the powerful concept of {\em SP-multiple-alignment} (SPMA, Appendix \ref{sp-multiple-alignment-appendix}), which may be seen as a generalisation of six other techniques for IC \cite[Section 5.7]{sp-micmup}, is bedrock in the workings of the SP System.

It is the concept of {\em SP-multiple-alignment} that is largely responsible for the AI-related strengths and potential of the SP System (Appendix \ref{sp-str-pot-appendix}), and for its several potential benefits and applications (Appendix \ref{benefits-and-applications-appendix}).

This strategy, aiming for SIABC, is unusual. And it is unusual to devote so much time and effort developing a theory. But that approach appears to be {\em essential} for the creation of a firm foundation for the development of human-like general intelligence or `artificial general intelligence' (AGI, Section \ref{top-down-strategy-section}).

\subsection{Demonstrations}

Most of the arguments are supported by one or more demonstrations with the SPCM. In some cases, relevant demonstrations have already been described in peer-reviewed papers so, to avoid unnecessary repetition, a summary of that earlier work is provided, with a citation.

\subsection{Source code and executable code for the SPCM}

Source code and Windows executable code for the SPCM may be downloaded as specified under the heading ``Code availability'' after the Conclusion (Section \ref{conclusion-section}).

\subsection{GPT-3}

The program GPT-3, developed by \href{https://openai.com/}{OpenAI} using DNN technologies, is attracting attention because of its impressive abilities to create what appears to be meaningful text.\footnote{See, for example, ``A robot wrote this entire article. Are you scared yet, human?'' (Guardian, 2020-09-08, \href{https://tinyurl.com/y3twe94t}{tinyurl.com/y3twe94t}).}

The main thrust of the criticism of GPT-3 by Gary Marcus and Ernest Davis,\footnote{``GPT-3, Bloviator: OpenAI's language generator has no idea what it's talking about'', MIT Technology Review, 2020-08-22, \href{tinyurl.com/yx8s25kh}{https://tinyurl.com/yx8s25kh}} is that it appears not to understand what it is writing, as evidenced by many errors described by Marcus and Davis in the kinds of commonsense reasoning that are straightforward for people.

Since the system relies on DNNs, it seems likely that it will also suffer from most of the problems described in Ford's book \cite{ford-2018}, and described in the main sections that follow.

\section{The need to bridge the divide between symbolic and non-symbolic kinds of knowledge and processing}\label{symbolic-non-symbolic-section}

This is the first of the sections mentioned above, each of which describes a significant problem in AI research, and how the SP System may solve it. Most of these sections, including this one, begin with one or more quotes from interviews  with Martin Ford of leading AI researchers, or observations by Ford himself (see {\em Architects of Intelligence}, \cite{ford-2018}).

\begin{quote}

    ``Some people still believe in symbolic AI, and they think there's potentially a need for a hybrid approach that incorporates both deep learning and more traditional approaches.'' Martin Ford  \cite[p.~84]{ford-2018}.

\end{quote}

\begin{quote}

    ``In my view, we need to bring together symbol manipulation, which has a strong history in AI, with deep learning. They have been treated separately for too long, and it's time to bring them together.'' Gary Marcus \cite[p.~318]{ford-2018}.

\end{quote}

\begin{quote}

    ``Many people will tell a story that in the early days of AI we thought intelligence was symbolic, but then we learned that was a terrible idea. It didn't work, because it was too brittle, couldn't handle noise and couldn't learn from experience. So we had to get statistical, and then we had to get neural. I think that's very much a false narrative. The early ideas that emphasize the power of symbolic reasoning and abstract languages expressed in formal systems were incredibly important and deeply right ideas. I think it's only now that we're in the position, as a field, and as a community, to try to understand how to bring together the best insights and the power of these different paradigms.'' Josh Tenenbaum \cite[pp.~476--477]{ford-2018}.

\end{quote}

With regard to the kinds of issues mentioned in the quotes above:

\begin{itemize}

    \item Given that people can and do learn and use symbolic systems like natural languages, mathematics, and logic, it is clear that our brains have the capacity to represent and to process those kinds of symbolic system.

    \item At the same time, any theory of AGI should be true to the kinds of things that can be done by: babies before they learn any symbolic system including natural language; or adults in performing putatively `non-symbolic' skills such as recognising things, playing tennis, making bread, and so on; or animals when they are engaged in such activities as foraging for edible plants, hunting prey, swinging from branch to branch through trees, and so on.

\end{itemize}

In the light of those two points, something is needed that bridges symbolic and non-symbolic kinds of knowledge and processing. The SP System provides a framework that is showing promise in those two areas:

\begin{itemize}

    \item The concept of {\em SP-symbol} in the SP System (Appendix \ref{outline-of-sp-system-appendix}) can represent a relatively large `symbolic' kind of thing such as a word, or it can represent a relatively fine-grained kind of thing such as a pixel in an image.

    \item The concept of {\em SP-pattern} (Appendix \ref{outline-of-sp-system-appendix}), with the concept of SPMA (Appendix \ref{sp-multiple-alignment-appendix}), provides a versatile framework for diverse aspects of intelligence (Appendix \ref{versatility-in-aspects of-intelligence-appendix}) including several forms of reasoning (Appendix \ref{versatility-in-reasoning-appendix}), and the representation of diverse kinds of knowledge (Appendix \ref{versatility-in-representation-of-knowledge-appendix}).

    \item The concept of SPMA also facilitates the seamless integration of diverse aspects of intelligence and diverse kinds of knowledge, in any combination (Appendix \ref{seamless-integration-appendix}), a kind of integration that appears to be essential in any system that aspires to AGI.

    \item As outlined in Appendix \ref{probabilistic-nature-of-sp-appendix}, the SP System is fundamentally probabilistic so in that respect it sits comfortably with the probabilistic nature of most of HLPC. But, when probabilities are at or near 0 or 1, the SP System has the potential to imitate the all-or-nothing nature of much of mathematics and logic \cite[Section 8]{sp-micmup}.

    \item The SP System has IC as a unifying principle, a principle which applies to both symbolic and non-symbolic kinds of knowledge.

    \item More specifically, the SP System conforms to the `DONSVIC' principle (the `Discovery of Natural Structures via Information Compression') \cite[Section 5.2]{sp-extended-overview}. This means that the system can discover relatively `symbolic' kinds of things such as words in otherwise non-symbolic information such as text without explicit markers for the beginnings and ends of words (as is the case with speech), and is also the case with text without punctuation or spaces between words ({\em ibid.}). It seems likely that similar principles apply to the discovery of `objects' via vision \cite[Sections 6.1 and 6.2]{sp-vision}.

\end{itemize}

\section{The tendency of deep neural networks to make large and unexpected errors in recognition}\label{dnns-easily-fooled-section}

\begin{quote}

    ``The vast majority of the dramatic advances we've seen over the past decade or so---everything from image and facial recognition, to language translation, to AlphaGo's conquest of the ancient game of Go---are powered by a technology known as deep learning, or deep neural networks.'' Martin Ford  \cite[p.~3]{ford-2018}.

\end{quote}

\begin{quote}

    ``In [a recent] paper \cite{brown-etal-2017}, [the authors] show how you can fool a deep learning system by adding a sticker to an image. They take a photo of a banana that is recognized with great confidence by a deep learning system and then add a sticker that looks like a psychedelic toaster next to the banana in the photo. Any human looking at it would say it was a banana with a funny looking sticker next to it, but the deep learning system immediately says, with great confidence, that it's now a picture of a toaster.'' Gary Marcus \cite[p.~318]{ford-2018}.

\end{quote}

Although DNNs often do well with images and speech (as described in the first quote above), they can make surprisingly big and unexpected errors in recognition (as described in the second quote). For example, a DNN may correctly recognise a picture of a car but may fail to recognise another slightly different picture of a car which, to a person, looks almost identical \cite{szegedy-etal-2014}. It has been reported that a DNN may assign an image with near certainty to a class of objects such as `guitar' or `penguin', when people judge the given image to be something like white noise on a TV screen or an abstract pattern containing nothing that resembles a guitar or a penguin or any other object \cite{nguyen-etal-2015}.

In a laboratory setting, errors like these may be interesting or even amusing. But to the extent that DNNs come into use in applications in, for example, commerce, administration, and defence, the consequences of errors can be expensive, or dangerous, or both those things:

\begin{quote}

    ``...~it is relatively easy to force [DNNs] to make mistakes that seem ridiculous, but with potentially catastrophic results. Recent tests have shown [how] autonomous vehicles could be made to ignore stop signs, and smart speakers could turn seemingly benign phrases into malware. ...~tiny changes to many of the pixels in an image could cause DNNs to change their decisions radically; a bright yellow school bus became, to the automated classifier, an ostrich. But the changes made were imperceptible to humans.'' \cite[p.~13]{edwards-2019}.

\end{quote}

Of course, there is potential for improvement in the performance of DNNs. But it seems that fixing problems in DNNs is difficult or impossible because of a fundamental weakness in them: that they represent and process knowledge in a way that is not transparent (Section \ref{transparency-in-rk-and-processing-section}). From an SP perspective, problems with DNNs arise because: 1) they do not have the tight focus on IC that is central in the workings of the SP System; and 2) because the basic organisation of DNNs has long been recognised as only a guess at how real nervous systems work.

From experience with the SP System to date, it seems very unlikely that the system would be vulnerable to the kinds of mistakes made by DNNs.

\subsection{Demonstrations}

Further evidence for the robustness of the SP framework in the recognition of patterns may be seen in:

\begin{itemize}

    \item The transparency of the SP System in both the representation and processing of knowledge (Section \ref{transparency-in-rk-and-processing-section}) means that all aspects of the workings of the SPCM are clear to see.

    \item In connection with the SPCM's strengths in generalisation (Section \ref{generalisation-section}):

    \begin{itemize}

        \item {\em Generalisation via unsupervised learning (Section \ref{generalisation-via-unsupervised-learning-section})}. There is evidence that the SP System can, via unsupervised learning, develop intuitively `correct' grammars for corresponding bodies of knowledge despite the existence of `dirty data' in the input data. Naturally, such grammars will have an impact on recognition in terms of those grammars.

        \item {\em Generalisation via perception (Section \ref{generalisation-via-perception-section})}. The SPCM demonstrates an ability to parse a sentence in a manner that is intuitively 'correct', despite errors of omission, commission, and substitution in the sentence that is to be parsed.

    \end{itemize}

\end{itemize}

\section{The need to strengthen the representation and processing of natural languages}\label{rk-processing-of-nl-section}

\begin{quote}

    ``...~I think that many of the conceptual building blocks needed for [artificial general intelligence] or human-like intelligence are already here. But there are some missing pieces. One of them is a clear approach to how natural language can be understood to produce knowledge structures upon which reasoning processes can operate.'' Stuart J. Russell \cite[p.~51]{ford-2018}.

\end{quote}

\begin{quote}

    ``...~a successful AI system needs some key abilities, including perception, vision, speech recognition, and action. These abilities help us to define artificial intelligence. We're talking about the ability to control robot manipulators, and everything that happens in robotics. We're talking about the ability to make decisions, to plan, and to problem-solve. We're talking about the ability to communicate, and so {\em natural language understanding also becomes extremely important to AI}.'' Stuart J. Russell \cite[p.~169]{ford-2018}, emphasis added.

\end{quote}

DNNs can do well in recognising speech. Also, they can produce impressive results in the translation of natural languages using a database of equivalences between surface structures that has been built up via human mark up and pattern matching, with English in some cases as a bridge between languages that are not English. And impressive results have been achieved with `zero-shot' learning\footnote{A zero-shot learning method aims to solve a task without receiving any example of that task at training phase.} \cite{radford-etal-2020}.

Despite successes like these, DNNs are otherwise weak in processing natural languages: they have no place for the kinds of syntactic structures that are recognised in theoretical linguistics, and which appear to be significant in human processing of language; they do not model the way in which people translate surface forms into meanings, and translate meanings into surface forms; and in such tasks as translation between languages, DNNs do not model the way in which, for people, meanings serve as an interlingua between different languages;

It seems likely that, without those kinds of abilities, AI systems will not achieve human levels of language understanding, or production, and it seems likely that AI systems will not reach the accuracy of translations between natural languages that can be achieved by human experts.

\subsection{Demonstrations}\label{demo-language-section}

The following subsections describe strengths of the SPCM in aspects of the processing of natural language. There are more demonstrations in \cite[Chapter 5]{wolff-2006} and \cite[Section 8]{sp-extended-overview}

\subsubsection{Parsing via SP-multiple-alignment}\label{parsing-via-spma-section}

As can be seen from the example in Figure \ref{parsing-kittens-figure} below, the SPMA construct in the SPCM lends itself well to the representation of syntactic knowledge and to its application in the parsing of a natural language sentence.

The New SP-pattern in row 0 shows a sentence to be parsed (`\texttt{t w o k i t t e n s p l a y}'), while the SP-patterns in rows 1 to 8, one SP-pattern per row, are Old SP-patterns, drawn from a relatively large repository of Old SP-patterns.

In this example, the Old SP-patterns represent grammatical categories including words. Thus the SP-pattern `\texttt{D Dp 4 t w o \#D}' represents the word `two' and the SP-symbols `\texttt{D}' and `\texttt{\#D}' show that it is a `determiner'. The SP-pattern `\texttt{NP D \#D N \#N \#NP}' represents a noun-phrase (`\texttt{NP ...~\#NP}') composed of a determiner (`\texttt{D \#D}') and a noun (`\texttt{N \#N}'), and so on.

How such an SP-multiple-alignment is created is described in Appendix \ref{sp-multiple-alignment-appendix}.


\begin{figure}[!htbp]
\fontsize{07.00pt}{08.40pt}
\centering
{\bf
\begin{BVerbatim}
0                          t w o              k i t t e n     s                    p l a y            0
                           | | |              | | | | | |     |                    | | | |
1                          | | |         Nr 5 k i t t e n #Nr |                    | | | |            1
                           | | |         |                 |  |                    | | | |
2                          | | |    N Np Nr               #Nr s #N                 | | | |            2
                           | | |    | |                         |                  | | | |
3                   D Dp 4 t w o #D | |                         |                  | | | |            3
                    | |          |  | |                         |                  | | | |
4            NP NPp D Dp         #D N Np                        #N #NP             | | | |            4
             |   |                                                  |              | | | |
5            |   |                                                  |         Vr 1 p l a y #Vr        5
             |   |                                                  |         |             |
6            |   |                                                  |  VP VPp Vr           #Vr #VP    6
             |   |                                                  |  |   |                    |
7 S Num    ; NP  |                                                 #NP VP  |                   #VP #S 7
     |     |     |                                                         |
8   Num PL ;    NPp                                                       VPp                         8
\end{BVerbatim}
}
\caption{The best SP-multiple-alignment created by the SPCM with a store of Old SP-patterns like those in rows 1 to 8 (representing syntactic structures, including words) and a New SP-pattern, `\texttt{t w o k i t t e n s p l a y}', shown in row 0, representing a sentence to be parsed. Adapted from Figure 1 in \cite{sp-intelligent-database}, with permission.}
\label{parsing-kittens-figure}
\end{figure}


\subsubsection{Discontinuous dependencies}\label{discontinuous-dependencies-section}

SPMAs may also represent and process discontinuous syntactic dependencies in natural language such as the dependency between the `number' (singular or plural) of the subject of a sentence, and the `number' (singular or plural) of the main verb, and that dependency may bridge arbitrarily large amounts of intervening structure (\cite[Section 8.1]{sp-extended-overview} and \cite[Section 5.4]{wolff-2006}).

In Figure \ref{parsing-kittens-figure}, a number dependency (plural) is marked by the SP-pattern in row 8: `\texttt{Num PL ; Np Vp}'. Here, the SP-Symbol `\texttt{Np}' marks the noun phrase as plural, and the SP-symbol `\texttt{Vp}' marks the verb phrase as plural.

In a similar way, the SPMA construct may mark the gender dependencies (masculine or feminine) within a sentence in French, and, within the same sentence, those two kinds of dependency may overlap without interfering with each other, as can be seen in \cite[Figure 5.8 in Section 5.4.1]{wolff-2006}.

\paragraph{Discontinuous dependencies in English auxiliary verbs.} The same method for encoding discontinous dependencies in natural language serves very well in encoding the intricate structure of such dependencies in English auxiliary verbs. This is described and demonstrated with the SPCM in \cite[Section 8.2]{sp-extended-overview} and \cite[Section 5.5]{wolff-2006}.

\subsubsection{Parsing which is Robust against errors of omission, commission, and substitution}

As can be seen in Figure \ref{generalisation-via-perception-figure}, and described in Section \ref{generalisation-via-perception-section}, the SP System has robust abilities to arrive at an intuitively `correct' parsing despite errors of omission, commission, and substitution, in the sentence being parsed. Naturally, there is a limit to how many errors can be tolerated.

\subsubsection{The representation and processing of semantic structures}

In case Figure \ref{generalisation-via-perception-figure} may have given the impression that the SPCM is only good for the processing of natural language syntax, it as well to emphasise that the system has strengths and potential in several other aspects of AI and the representation of knowledge (summarised in Appendix \ref{sp-str-pot-appendix}), including:

\begin{itemize}

    \item Several kinds of probabilistic reasoning (Section \ref{probabilistic-reasoning-section}, \cite[Section 10]{sp-extended-overview}, \cite[Chapter 7]{wolff-2006}).

    \item The representation and processing of:

    \begin{itemize}

        \item Ontologies (\cite{sp-ontologies}, \cite[Section 13.4.3]{wolff-2006}).

        \item Class hierarchies with inheritance of attributes, including  cross-classification (multiple inheritance) \cite[Section 6.4]{wolff-2006}).

        \item Part-whole hierarchies and their integration with class-inclusion hierarchies (\cite[Section 9.1]{sp-extended-overview}, \cite[Section 6.4]{wolff-2006}).

        \item Decision networks and trees \cite[Section 7.5]{wolff-2006}.

        \item Relational tuples (\cite{sp-intelligent-database}, \cite[Section 13.4.6.1]{wolff-2006}).

        \item If-then rules \cite[Section 7.6]{wolff-2006}.

        \item Associations of medical signs and symptoms \cite{sp-medical-diagnosis}.

        \item Causal relations \cite[Section 7.9]{wolff-2006}.

        \item And concepts in mathematics and logic, such as `function', `variable', `value', `set' and `type definition' \cite[Chapter 10]{wolff-2006}.

    \end{itemize}

\end{itemize}

The use of one simple framework for the representation and processing of knowledge facilitates the seamless integration of different kinds of knowledge, in any combination.

Any or all of such structures or processes may function as the meanings of natural language.

\subsubsection{The integration of syntax and semantics}\label{integration-syntax-semantics-section}

Because of the versatility of SP-patterns with SPMAs (Appendix \ref{sp-str-pot-appendix}), there is one framework which lends itself to the representation and processing of both the syntax and semantics of natural languages. It is likely that this will facilitate the integration of syntax with semantics, so that surface forms may be translated into meanings, and {\em vice versa}. This is demonstrated in \cite[Section 5.7]{wolff-2006}

\subsubsection{One mechanism for both the parsing and production of natural language}

A neat feature of the SP System is that the production of natural language may be achieved by the application of IC, using {\em exactly} the same mechanisms as are used for the parsing or understanding of natural language. An explanation of how this is possible is given in \cite[Section 4.5]{sp-extended-overview}) and \cite[Section 5.7.1]{wolff-2006}.

\section{The challenges of unsupervised learning}\label{challenge-of-unsupervised-learning-section}

\begin{quote}

    ``Unsupervised learning represents one of the most promising avenues for progress in AI. ...~However, it is also one of the most difficult challenges facing the field. {\em A breakthrough that allowed machines to efficiently learn in a truly unsupervised way would likely be considered one of the biggest events in AI so far, and an important waypoint on the road to AGI.}'' Martin Ford \cite[pp.~11--12]{ford-2018}, emphasis added.

\end{quote}

\begin{quote}

    \sloppy ``Until we figure out how to do this unsupervised/self-supervised/predictive learning, we're not going to make significant progress because I think that's the key to learning enough background knowledge about the world so that common sense will emerge.'' Yann Lecun \cite[p.~130]{ford-2018}.

\end{quote}

\begin{quote}

    ``Unsupervised learning is hugely important, and we're working on that.'' Demis Hassabis \cite[p.~170]{ford-2018}.

\end{quote}

From these remarks it can be seen that the development of unsupervised learning is regarded as an important challenge for AI research today. So it should be of interest that unsupervised learning in the SP System is founded on an earlier programme of research on the unsupervised learning of language \cite{wolff-1988}, that it is a key part of the SP System now and in developments envisaged for the future \cite[Section 12]{sp-palade-wolff}, and that the SPCM has already demonstrated relevant strengths (more below).

In the SP programme of research, unsupervised grammatical inference is regarded as a paradigm or framework for other kinds of unsupervised learning, not merely the learning of syntax. In addition to the learning of syntactic structures \cite[Section 12.3]{sp-palade-wolff}, it may, for example, provide a model for the unsupervised learning of non-syntactic semantic structures \cite[Section 12.1]{sp-palade-wolff}, and for learning the integration of syntax with semantics \cite[Section 12.3]{sp-palade-wolff}.

With further development, unsupervised learning in the SPCM may itself be a good foundation for `supervised' kinds of learning, such as learning by being told, learning by imitation, learning via rewards and punishments, and so on.

\subsection{Demonstrations}

There is a detailed description of unsupervised learning in the SP System in \cite[Chapter 9]{wolff-2006} and a briefer account in \cite[Section 5]{sp-extended-overview}.

Here is a simple example of how the SPCM may achieve unsupervised learning. Figure \ref{unsupervised-learning-figure} (a) shows a set of New SP-patterns supplied to the SPCM as data for learning, and Figure \ref{unsupervised-learning-figure} (b) shows the best of several alternative {\em SP-grammars} created by the SPCM, where an `SP-grammar' is simply a set of Old SP-patterns. The SP-grammar which is `best' is the one with the smallest value for $T$, where $T = G + E$, $G$ is the size of the SP-grammar in bits, and $E$ is the size of the New SP-patterns when they have been encoded in terms of the SP-grammar (as described in \cite[Section 5.1.2]{sp-extended-overview} and \cite[Section 9.2.3]{wolff-2006}).

\begin{figure}[!htbp]
\fontsize{10.00pt}{12.00pt}
\centering
{\bf
\begin{BVerbatim}
j o h n r u n s
m a r y r u n s
j o h n w a l k s
m a r y w a l k s

(a)

< %2 1 s >
< %3 2 m a r y >
< %3 3 j o h n >
< %1 4 r u n >
< %1 5 w a l k >
< 6 < %3 > < %1 > < %2 > >

(b)
\end{BVerbatim}
}
\caption{(a) A set of New SP-patterns supplied to the SPCM as data for unsupervised learning. (b) The best of several alternative SP-grammars created by the program, where the meaning of `best' is described in the text.}
\label{unsupervised-learning-figure}
\end{figure}

The SP-grammar in the figure can be interpreted quite simply like this: The last SP-pattern, with the identifier `\texttt{6}', describes how a sentence may be constructed from SP-patterns in classes `\texttt{\%3}', followed by `\texttt{\%1}', followed by `\texttt{\%2}'; SP-patterns of class `\texttt{\%3}' are `\texttt{(< \%3 2 m a r y >)}' and `\texttt{(< \%3 3 j o h n >)}'; SP-patterns of class `\texttt{\%1}' are `\texttt{(< \%1 4 r u n >)}' and `\texttt{(< \%1 5 w a l k >)}'; while `\texttt{\%2}' identifies the singleton `\texttt{(< \%2 1 s >)}'.

Any such SP-grammar may serve in the parsing of an appropriate sentence (as in Figure \ref{parsing-kittens-figure}).

With appropriate input, the SPCM can also produce plausible SP-grammars for sentences like `\texttt{(t h a t b o y r u n s)}', `\texttt{(t h a t g i r l r u n s)}', and so on. But the program has limitations described in Appendix \ref{shortcomings-sp-learning-appendix}: that it cannot yet identify structures like clauses and phrases at intermediate levels between words and sentences; and it cannot yet learn discontinuous dependencies of the kind described in Section \ref{discontinuous-dependencies-section}.

It appears that such problems are soluble and that their solution would greatly enhance the capabilities of the SPCM in unsupervised learning.

\section{The need for a coherent account of generalisation, under-generalisation, and over-generalisation}\label{generalisation-section}

\begin{quote}

    ``Many of us think that we are ...~missing the basic ingredients needed [for generalization], such as the ability to understand causal relationships in data---an ability that actually enables us to generalize and to come up with the right answers in settings that are very different from those we've been trained in.'' Yoshua Bengio \cite[p.~18]{ford-2018}.

\end{quote}

\begin{quote}

    ``...~we might have a photograph, where we've got all the pixels in the image, and then we have a label saying that this is a photograph of a boat, or of a Dalmatian dog, or of a bowl of cherries. In supervised learning for this task, the goal is to find a predictor, or a hypothesis, for how to classify images in general.'' Stuart J. Russell \cite[p.~41]{ford-2018}.

\end{quote}

\begin{quote}

    ``The theory [worked on by Roger Shepard and Joshua Tenenbaum] was of how humans, and many other organisms, solve the basic problem of generalization, which turned out to be an incredibly deep problem. ...~The basic problem is, how do we go beyond specific experiences to general truths? Or from the past to the future?'' Joshua Tenenbaum \cite[p.~468]{ford-2018}.

\end{quote}

An important issue in unsupervised learning is how to generalise `correctly' from the specific information which provides the basis for learning, without over-generalisation (`under-fitting') or under-generalisation (`over-fitting'). This issue is discussed quite fully in \cite[Section 5.3]{sp-extended-overview} (see also \cite[Section 9.5.3]{wolff-2006} and \cite[Section V-H]{sp-alternatives}). Because it is an important issue, the main elements of the solution proposed in the SP Theory of Intelligence are described here.

In the SP Theory of Intelligence, generalisation may be seen to occur in two aspects of AI: as part of the process of unsupervised learning; and as part of the process of recognition. Those two aspects are considered in the following two subsections.

\subsection{Generalisation via unsupervised learning}\label{generalisation-via-unsupervised-learning-section}

\begin{figure}[!htbp]
\centering
\includegraphics[width=0.5\textwidth]{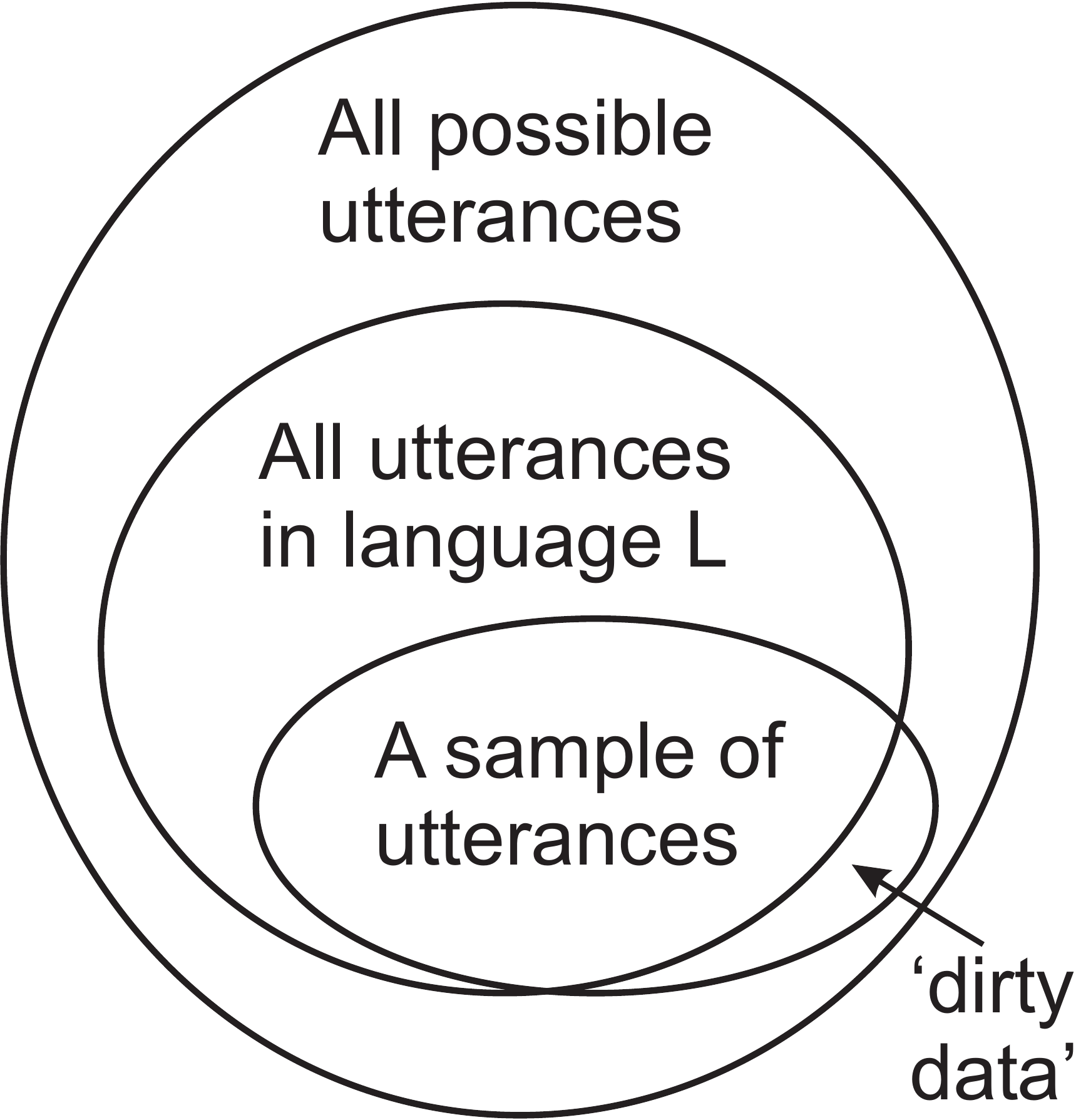}
\caption{Categories of utterances involved in the learning of a first language, {\bf L}. In ascending order of size, they are: the finite sample of utterances from which a child learns; the (infinite) set of utterances in {\bf L}; and the larger (infinite) set of all possible utterances. Adapted from Figure 7.1 in \protect\cite{wolff-1988}, with permission.}
\label{generalisation-learning-figure}
\end{figure}

The generalisation issue arises quite clearly in considering how children learn their native language or languages, as illustrated in Figure \ref{generalisation-learning-figure}.

Each child learns a language {\bf L} from a sample of things that they hear being said by people around them. Although the sample is normally large, it is nevertheless, finite.\footnote{It is assumed here that the learning we are considering is unsupervised. This is because of evidence that, although correction of errors by adults may be helpful, children are capable of learning a first language without that kind of error correction (\cite{lenneberg-1962,brown-2014}).} That finite sample is shown as the smallest envelope in Figure \ref{generalisation-learning-figure}.

The variety of `legal' utterances in the language {\bf L} is represented by the next largest envelope in the figure. The largest envelope represents the variety possible utterances, both those in {\bf L} and everything else including grunts, gurgles, false starts, and so on.

Each of the two larger envelopes represents a set that is infinite in size but, in accordance with principles pioneered by Georg Cantor (see, for example, \cite{dauben-1990}), the set of all possible utterances is larger than the set of utterances in {\bf L}. This is much like the way that the set of all integers is larger than the set of even integers, or the set of odd integers, although each of those sets is infinite in size.

The difference in size between the smallest envelope and the middle-sized envelope represents correct generalisations. If a learning system creates a grammar that generates {\bf L} plus some other utterances, then it over-generalises, and if the grammar generates a subset of the utterances in {\bf L}, then it under-generalises.

An interesting feature of the learning of a first language by children is that their finite sample of what people are saying includes some things that are {\em not} in {\bf L} (because of slips of the tongue and the like) as well as many things that are in {\bf L}. In the figure, the `illegal' utterances that children hear are marked as `dirty data'.

Challenges in understanding how young children learn their first language (or languages) are:

\begin{itemize}

    \item How they generalise correctly without over-generalisation. In this connection, it is interesting that young children often over-generalise, saying things like `mouses' instead of `mice' (applying a pluralisation rule too widely), or saying things like `hitted' instead of `hit' (applying a past tense rule too widely). But they normally weed out such over-generalisations when they are older.

    \item Young children may also under-generalise, but, because the children are young, such under-generalisations are difficult to distinguish from the fact that there is much of the language that they don't yet know.

    \item As with over-generalisations, it seems that children eventually learn {\bf L} without their learning being distorted or corrupted by dirty data. Anything that is `wrong' that they do learn would probably be regarded as dialect rather than errors in learning.

\end{itemize}

Judging by the quotations at the beginning of this section, and elsewhere in \cite{ford-2018}, and judging by other writings about language learning and other kinds of learning, there is no coherent theory of generalisation, and over- or under-generalisation, that is widely recognised by researchers in AI or psychology.

For reasons that would take us too far afield to explain, `nativist' theories of first language learning, such as that proposed by Noam Chomsky \cite{chomsky-1965} and others, will not suffice.

What follows is a summary of what is already largely incorporated in the SPCM:

\begin{enumerate}

    \item Unsupervised learning in the SP Theory of Intelligence may be seen as a process of compressing a body of information, {\bf I}, to achieve lossless compression of {\bf I} into a structure {\bf T}, where the size of {\bf T} (represented by $t$) is as small as can be achieved with the available computational resources.

    \item {\bf T} may be divided into two parts: a {\em grammar}, {\bf G} of size $g$; and an {\em encoding}, {\bf E}, of {\bf I} in terms of {\bf G}, where the size of {\bf E} is $e$. Clearly, $t = g + e$.

    \item Discard {\bf E} and retain {\bf G}.

    \item Provided the compression of {\bf I} has been done quite thoroughly, {\bf G} may be seen to be a theory of {\bf I} which generalises `correctly' beyond {\bf I} without either over- or under-generalisations.

\end{enumerate}

Why should we have more confidence in {\bf G} as a source of `correct' generalisations than anything else? The best answer at present seems to be that {\bf G} may be seen as a distillation of redundancies in {\bf I}, whereas {\bf E} is largely the non-redundant aspects of {\bf I}. Since redundancy equates largely with repetition, and since repetition provides the basis for inductive inference, {\bf G} may be seen to be the most promising source of generalisations. Each typo, cough, or similar kind of dirty data is normally rare in a given context and will normally be recorded in {\bf E}, not {\bf G}---and so {\bf E} may be discarded.

Much of this thinking derives from \cite[Section 3.4]{solomonoff-1997}:

\begin{verbatim}

    ``Given a set of positive cases of acceptable sentences and several grammars, any of which is able to generate all of the sentences, what goodness of fit criterion should be used? It is clear that the `ad-hoc grammar,' that lists all of the sentences in the corpus, fits perfectly. The `promiscuous grammar' that accepts any conceivable sentence, also fits perfectly. The first grammar has a long description; the second has a short description. It seemed that some grammar half-way between these, was `correct'---but what criterion should be used?''

\end{verbatim}

\subsubsection{Demonstrations}

Informal tests with the SPCM \cite[Chapter 9]{wolff-2006}, suggest that it may learn what are intuitively `correct' structures, in spite of being supplied with data that is incomplete in the sense that generalisations are needed to produce a `correct' result. Here is a simple example:

\begin{itemize}

    \item The SPCM has been run with a set of New SP-patterns like those shown in Figure \ref{unsupervised-learning-figure} (a) but {\em without} the last New SP-pattern `\texttt{(m a r y w a l k s)}'.

    \item Despite the omission of that last New SP-pattern, the best SP grammar that is created by the SPCM is {\em exactly} the same as before (shown in Figure \ref{unsupervised-learning-figure} (b)).

    \item In effect the SPCM has made an inference or prediction: that although the sentence `\texttt{(m a r y w a l k s)}' was not amongst the New SP-patterns that the program received as data, it has in effect made the prediction that `\texttt{(m a r y w a l k s)}' is a `legal' sentence.

\end{itemize}

Similar results have been obtained with incomplete sets of sentences like `\texttt{(t h a t b o y r u n s)}', `\texttt{(t h a t g i r l r u n s)}', and so on.

Of course, simple examples like these are only a beginning, and it will be interesting to see how the SPCM generalises with more ambitious examples, especially when the program has reached the stage when it can produce plausible SP-grammars with natural language.

Even now, there is a reason to have confidence in the model of generalisation that has been described: because it's basis in compression of information is consonant with much other evidence for the significance of IC in the workings of brains and nervous systems \cite{sp-compression}.

{\em Dirty data}. With regard to `dirty data', mentioned above and shown in Figure \ref{generalisation-learning-figure}: in informal experiments with models of language learning developed in earlier research that have IC as a unifying principle: ``In  practice,  the  programs  MK10  and  SNPR  have  been  found  [to produce intuitively `correct' results but] to  be  quite  insensitive  to  errors (of omission, addition, or substitution) in their data.'' \cite[p.~209]{wolff-1988}. In that connection, it is suggested that:

\begin{quote}

    ``A strength of the theory is that it neatly explains how children can learn from [dirty] data without being thrown off by errors: any  particular  error  is,  by  its  nature,  rare  and  so  in  the  search  for  useful  (common)  structures, it is discarded along with many other candidate structures. (If an error is not rare it is likely to acquire the status of a dialect or idiolect variation and cease to be regarded as an error.)'' \cite[p.~208]{wolff-1988}

\end{quote}

\subsection{Generalisation via perception}\label{generalisation-via-perception-section}

The SPCM has a robust ability to recognise things or to parse natural language despite errors of omission, commission, or substitution in what is being recognised or parsed. Incidentally, it is assumed here that recognition in any sensory modality may be understood largely as parsing, as described in \cite[Section 4]{sp-vision}.

\subsubsection{Demonstration}\label{errors-paqsing-section}

The example here makes reference, first, to Figure \ref{parsing-kittens-figure} in Section \ref{demo-language-section}, which shows how an SPMA may achieve the effect of parsing the sentence `\texttt{t w o k i t t e n s p l a y}' in terms of grammatical categories, including words.

\begin{figure}[!htbp]
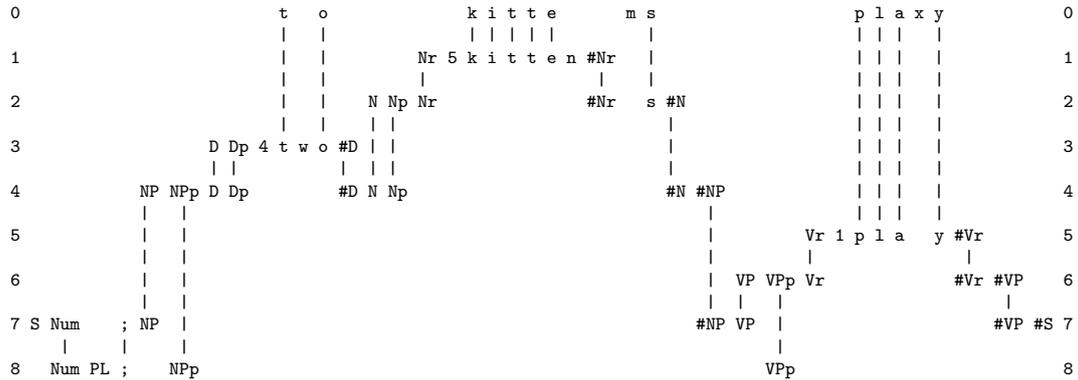

\fontsize{07.00pt}{08.40pt}
\centering
{\bf
\begin{BVerbatim}
0                          t   o              k i t t e       m s                    p l a x y            0
                           |   |              | | | | |         |                    | | |   |
1                          |   |         Nr 5 k i t t e n #Nr   |                    | | |   |            1
                           |   |         |                 |    |                    | | |   |
2                          |   |    N Np Nr               #Nr   s #N                 | | |   |            2
                           |   |    | |                           |                  | | |   |
3                   D Dp 4 t w o #D | |                           |                  | | |   |            3
                    | |          |  | |                           |                  | | |   |
4            NP NPp D Dp         #D N Np                          #N #NP             | | |   |            4
             |   |                                                    |              | | |   |
5            |   |                                                    |         Vr 1 p l a   y #Vr        5
             |   |                                                    |         |               |
6            |   |                                                    |  VP VPp Vr             #Vr #VP    6
             |   |                                                    |  |   |                      |
7 S Num    ; NP  |                                                   #NP VP  |                     #VP #S 7
     |     |     |                                                           |
8   Num PL ;    NPp                                                         VPp                           8

\end{BVerbatim}
}
\caption{As in Figure \ref{parsing-kittens-figure} but with errors in the sentence in row 0 (`\texttt{t o k i t t e m s p l a x y}'): an error of omission (`\texttt{t o}' instead of `\texttt{t w o}'), an error of substitution (`\texttt{k i t t e m s}' instead of `\texttt{k i t t e n s}'), and an error of addition (`\texttt{p l a x y}' instead of `\texttt{p l a y}'). Adapted from Figure 2 in \cite{sp-intelligent-database}, with permission.}
\label{generalisation-via-perception-figure}
\end{figure}

To illustrate generalisation via perception, Figure \ref{generalisation-via-perception-figure}, below, shows how the SPCM, with the New SP-pattern with errors, `\texttt{t o k i t t e m s p l a x y}', may achieve what is intuitively a `correct' analysis of the sentence despite the errors, which are described in the caption of the figure.


This kind of recognition in the face of errors may be seen as a kind of generalisation, where an incorrect form is generalised to the correct form.

There is relevant discussion in \cite[Sections 4.1 and 4.2]{sp-vision}.

\section{How to learn usable knowledge from a single exposure or experience}\label{learning-from-a-single-experience-section}

\begin{quote}

    ``How do humans learn concepts not from hundreds or thousands of examples, as machine learning systems have always been built for, but from just one example? ...~Children can often learn a new word from seeing just one example of that word used in the right context, ...~You can show a young child their first giraffe, and now they know what a giraffe looks like; you can show them a new gesture or dance move, or how you use a new tool, and right away they’ve got it ...'' Joshua Tenenbaum \cite[p.~471]{ford-2018}.

\end{quote}

\begin{quote}

    ``There's also `zero-shot learning,' where people are trying to build programs that can learn when they see something even for the first time. And there is `one-shot learning' where a program sees a single example, and they're able to do things.'' Oren Etzioni \cite[p.~500]{ford-2018}.

\end{quote}

Most DNNs incorporate some variant of the idea that, in learning, neural connections are gradually strengthened or weakened, a concept of learning which often draws its inspiration, directly or indirectly, from Donald Hebb's concept of learning:

\begin{quote}

``When an axon of cell A is near enough to excite a cell B and repeatedly or persistently takes part in firing it, some growth process or metabolic change takes place in one or both cells such that A's efficiency, as one of the cells firing B, is increased.'' \cite[Location 1496]{hebb-1949}.

\end{quote}

\noindent or, more briefly, ``Neurons that fire together, wire together.''

Although this seems to reflect the way that it takes time to learn a complex skill such as playing the piano well, or competition-winning abilities in pool, billiards, or snooker, this feature of DNNs conflicts with the undoubted fact that people can and often do learn usable knowledge from a single occurrence or experience: memories for significant events that we experience only once may be retained for many years; in everyday activities such as having a conversation, we can quickly assimilate what is said to us and retain it for minutes, days or years.

In the SP System, any of the newly-received knowledge from the system's environment may serve immediately in any kind of thinking or activity. As an analogy, if we ask the way from someone, the guidance that we get can be put into effect at once, without the need for any kind of gradual strengthening of links. If a child touches something hot, he or she is likely to retain what they have learned for the rest of their lives, without the need for repetition. We may remember winning big on the lottery without the need for repeated experiences to drum it in (welcome as that may be!).

It is true that DNNs require the taking in of information supplied by the user, and may thus be said to have learned something from a single exposure or experience. But unlike the SP System, that knowledge cannot be used immediately, as illustrated by the parsing of a New sentence (fresh from the system's environment) in the parsing shown in Figure \ref{parsing-kittens-figure}. By contrast, any information taken in by a DNN at the beginning of a training session does not become usable until much more information has been taken in.

\subsection{Slow learning of complex knowledge or skills}

By contrast with the fast learning of simple things, the SP System also provides an explanation for why people are relatively slow at learning complex things like how to play the piano well, or how to be an expert medical doctor. It is likely that the slow learning of these kinds of things is partly because there is a lot to be learned, and partly because that kind of learning requires a time-consuming search through a large abstract space of ways in which the knowledge may be structured in order to compress it and thus arrive at an efficient configuration.

\subsection{Demonstration}

Figure \ref{parsing-kittens-figure} (Section \ref{demo-language-section}) provides a good example of one-shot learning. The sentence to be parsed, `\texttt{t w o k i t t e n s p l a y}' is read in directly as a New SP-pattern from the SP System's environment, and is thus learned. Immediately, it may play a part in the process of parsing, and the same would be true for New SP-patterns in any other kind of operation such as reasoning, problem solving, planning or the like. There is no need for any slow process of strengthening links across the several layers of a DNN.

This accords with our experience in any ordinary conversation amongst two or more people. Each person may respond immediately to what another person has said, or at least within a few seconds if `deep' matters are being discussed. And, and as noted earlier, we can normally remember most of what other people have said for minutes, days, or longer.

\section{How to achieve transfer learning (incorporating old knowledge in new)}\label{transfer-learning-section}

\begin{quote}

    ``Transfer learning is where you usefully transfer your knowledge from one domain to a new domain that you've never seen before, it's something humans are amazing at. If you give me a new task, I won't be terrible at it out of the box because I'll bring some knowledge from similar things or structural things, and I can start dealing with it straight away. That's something that computer systems are pretty terrible at because they require lots of data and they're very inefficient.'' Demis Hassabis \cite[p.~174]{ford-2018}.

\end{quote}

\begin{quote}

    ``Humans can learn from much less data because we engage in transfer learning, using learning from situations which may be fairly different from what we are trying to learn.'' Ray Kurzweil \cite[p.~230]{ford-2018}.

\end{quote}

\begin{quote}

    ``We need to figure out how to think about problems like transfer learning, because one of the things that humans do extraordinarily well is being able to learn something, over here, and then to be able to apply that learning in totally new environments or on a previously unencountered problem, over there.'' James Manyika \cite[p.~276]{ford-2018}.

\end{quote}

Transfer learning---meaning the use of old learning to facilitate later learning---is fundamental in the SP System.

Because the system does not suffer from catastrophic forgetting (Section \ref{catastrophic-forgetting-section}), and because one-shot learning is possible and normal (Section \ref{learning-from-a-single-experience-section}), SP-patterns that have been learned at any stage, will be available immediately in the system's repository of Old SP-patterns for use later. Some of the several ways in which previously-learned Old SP patterns may help with transfer learning is illustrated in the examples described here:

\begin{itemize}

    \item {\em Partial matching in unsupervised learning}. Section \ref{demo-transfer-learning-section} describes how the SPCM may achieve transfer learning via partial matching between two SP-patterns.

    \item {\em Interpretation of New information in terms of pre-existing Old information}. The SPMA in Figure \ref{parsing-kittens-figure} shows how a New SP-pattern (\texttt{t w o k i t t e n s p l a y}) may be analysed or interpreted in terms of pre-existing Old SP-patterns. Within the processes for unsupervised learning (Appendix \ref{sp-unsupervised-learning-appendix} and Section \ref{challenge-of-unsupervised-learning-section}), this kind of analysis yields an encoding as described in \cite[Section 4.1]{sp-extended-overview} and \cite[Section 3.5]{wolff-2006}. This encoding may then itself feed into further processes for unsupervised learning.

    \item {\em Building syntactic and semantic knowledge}. In unsupervised learning in the SPCM (\cite[Section 5]{sp-extended-overview}, \cite[Chapter 9]{wolff-2006}), knowledge about words, which is acquired relatively early, is incorporated later into sentence structures. It is envisaged that similar processes will apply in the learning of non-syntactic semantic structures, and in the marriage of syntax with semantics.

\end{itemize}

\subsection{Demonstration}\label{demo-transfer-learning-section}

Here is a simple example of how the SPCM may achieve transfer learning between one Old SP-pattern and one New SP-pattern:

\begin{itemize}

    \item At the beginning, there is one Old SP-pattern in store: `\texttt{< \%1 3 t h a t b o y r u n s >}'.

    \item Then a New SP-pattern is received: `\texttt{t h a t g i r l r u n s}'.

    \item The best SP-multiple alignment for these two SP-patterns is shown in Figure \ref{transfer-learning-alignment-figure}.

    \item From that SP-multiple-alignment, the SPCM derives SP-patterns as shown in Figure \ref{transfer-learning-grammar-figure}.

    \item Because IC in the SPCM is always lossless (Appendix \ref{ic-via-icmup-appendix}), the SP-grammar in Figure \ref{transfer-learning-grammar-figure} is equivalent to the original two SP-patterns from which it is derived.

\end{itemize}

\begin{figure}[!htbp]
\fontsize{10.00pt}{12.00pt}
\centering
{\bf
\begin{BVerbatim}
0        t h a t g i r l r u n s   0
         | | | |         | | | |
1 < %1 3 t h a t b o y   r u n s > 1
\end{BVerbatim}
}
\caption{The best SPMA created by the SPCM between the Old SP-pattern `\texttt{< \%1 3 t h a t b o y r u n s >}' and the New SP-pattern `\texttt{t h a t g i r l r u n s}'.}
\label{transfer-learning-alignment-figure}
\end{figure}

\begin{figure}[!htbp]
\fontsize{10.00pt}{12.00pt}
\centering
{\bf
\begin{BVerbatim}
< %1 1 t h a t >
< %2 2 r u n s >
< %3 3 b o y >
< %3 4 g i r l >
< 5 < %1 > < %3 > < %2 > >
\end{BVerbatim}
}
\caption{The SP-grammar created by the SPCM from the SPMA shown in Figure \ref{transfer-learning-alignment-figure}.}
\label{transfer-learning-grammar-figure}
\end{figure}

\section{How to increase the speed of learning in AI systems, and how to reduce the demands of AI learning for large volumes of data, and for large computational resources}\label{comp-resources-speed-volumes-of-data-section}

\begin{quote}

    ``People can learn from very few examples and generalize. We don't know how to build machines that can do that.'' Cynthia Breazeal \cite[p.~456]{ford-2018}.

\end{quote}

\begin{quote}

    ``We can imagine systems that can learn by themselves without the need for huge volumes of labeled training data.'' Martin Ford \cite[p.~12]{ford-2018}.

\end{quote}

\begin{quote}

    ``...~the first time you train a convolutional network you train it with thousands, possibly even millions of images of various categories.'' Yann LeCun \cite[p.~124]{ford-2018}.

\end{quote}

\begin{quote}

    ``[A] stepping stone [towards artificial general intelligence] is that it's very important that [AI] systems be a lot more data-efficient. So, how many examples do you need to learn from? If you have an AI program that can really learn from a single example, that feels meaningful. For example, I can show you a new object, and you look at it, you're going to hold it in your hand, and you're thinking, `I've got it.' Now, I can show you lots of different pictures of that object, or different versions of that object in different lighting conditions, partially obscured by something, and you'd still be able to say, `Yep, that's the same object.' But machines can't do that off of a single example yet. That would be a real stepping stone to [artificial general intelligence] for me.'' Oren Etzioni \cite[p.~502]{ford-2018}.

\end{quote}

In connection with the large volumes of data and large computational resources that are often associated with the training of DNNs, it has been discovered by Emma Strubell and colleagues \cite{strubell-etal-2019} that the process of training a large AI model can emit more than 626,000 pounds of carbon dioxide, which is equivalent to nearly five times the lifetime emissions of the average American car, including the manufacture of the car itself.

Issues mentioned in the quotes above are considered in subsections that follow. They are: (1) The relatively slow speeds of learning with DNNs, (2) The relatively large volumes of data required by DNNs, and (3) The large computational resources that are often associated with the training of DNNs. These things seem to conflict with how people can learn many kinds of things fast, with relatively little data, and with a brain that runs on less than 20 watts \cite{jabr-2012}.

\subsection{Speeds of learning}\label{speeds-of-learning-section}

The SP System suggests two main ways in which gains in speed may be achieved:

\begin{itemize}

    \item {\em Learning via a single exposure or experience}. Take advantage of the way in which the SPCM may learn usable knowledge from a single exposure or experience (Section \ref{learning-from-a-single-experience-section}).

    \item {\em Transfer learning}. Take advantage of the way in which the SPCM may incorporate already-stored knowledge in the learning of something new (Section \ref{transfer-learning-section}).

\end{itemize}

What about the slow learning of complex knowledge or skills such as the learning of a natural language or learning to play the piano well? In terms of the SP System, it seems likely that slow learning in these cases can be explained by the need to search a very large space of alternative structures. Although incremental, heuristic search can avoid the need for comprehensive search, there is still a lot of processing to be done.

In connection with these issues, a point which deserves emphasis is that unsupervised learning in the SP System is entirely different from Hebbian learning (Appendix \ref{sp-unsupervised-learning-appendix}). While the SP System can explain fast learning in some situations and slow learning in others, Hebbian learning insists, contrary to biological fact, that all kinds of learning have to be slow.

\subsection{The relatively large volumes of data required by DNNs}\label{volumes-of-data-section}

As with speeds of learning (Section \ref{speeds-of-learning-section}), the SPCM may reduce the amount of data needed for learning via: 1) Learning via a single exposure or experience (Section \ref{learning-from-a-single-experience-section}); and 2) Transfer learning, making use of Old knowledge in learning from New data (Section \ref{transfer-learning-section}).

With regard to the second of these features, a well-developed SP Machine (Appendix \ref{unfinished-business-appendix}) may be needed to create useful SP-grammars if the volumes of data to be processed are large. But once a good SP-grammar has been created, it is likely that the use of the SP-grammar for the encoding of new data may be done with less powerful machines.

\subsection{The large computational resources that are often associated with the training of DNNs}

As with speeds of learning (Section \ref{speeds-of-learning-section}) and volumes of data (Section \ref{volumes-of-data-section}), computational resources for learning may be reduced via one-shot learning with transfer learning. Again, a relatively powerful SP Machine may be needed to create one or more good SP-grammars for each kind of data, but once such SP-grammars have been developed, the process of encoding new data in terms of an SP-grammar may be done with less powerful machines.

\section{The need for transparency in the representation and processing of knowledge}\label{transparency-in-rk-and-processing-section}

\begin{quote}

    ``...~if regulation is intended to think about questions of safety, questions of privacy, questions of transparency, questions around the wide availability of these techniques so that everybody can benefit from them---then I think those are the right things that AI regulation should be thinking about.'' James Manyika \cite[p.~283]{ford-2018}.

\end{quote}

\begin{quote}

    ``Although Bayesian updating is one of the major components in machine learning today, there has been a shift from Bayesian networks to deep learning, which is less transparent.'' Judea Pearl \cite[p.~363]{ford-2018}.

\end{quote}

\begin{quote}

    ``The current machine learning concentration on deep learning and its non-transparent structures is such a hang-up.'' Judea Pearl \cite[p.~369]{ford-2018}.

\end{quote}

It is now widely recognised that a major problem with DNNs is that the way in which learned knowledge is represented in such systems is far from being comprehensible by people, and that the way in which DNNs arrive at their conclusions is difficult for people to understand. These deficiencies are of concern for reasons of safety, legal liability, fixing problems in systems that use DNNs, and perhaps more.\footnote{See, for example, ``Inside DARPA's push to make artificial intelligence explain itself'', {\em The Wall Street Journal}, 2017-08-10, \href{https://tinyurl.com/y9vrov2k}{tinyurl.com/y9vrov2k}.}

\begin{figure}[!htbp]
\centering
\includegraphics[width=0.9\textwidth]{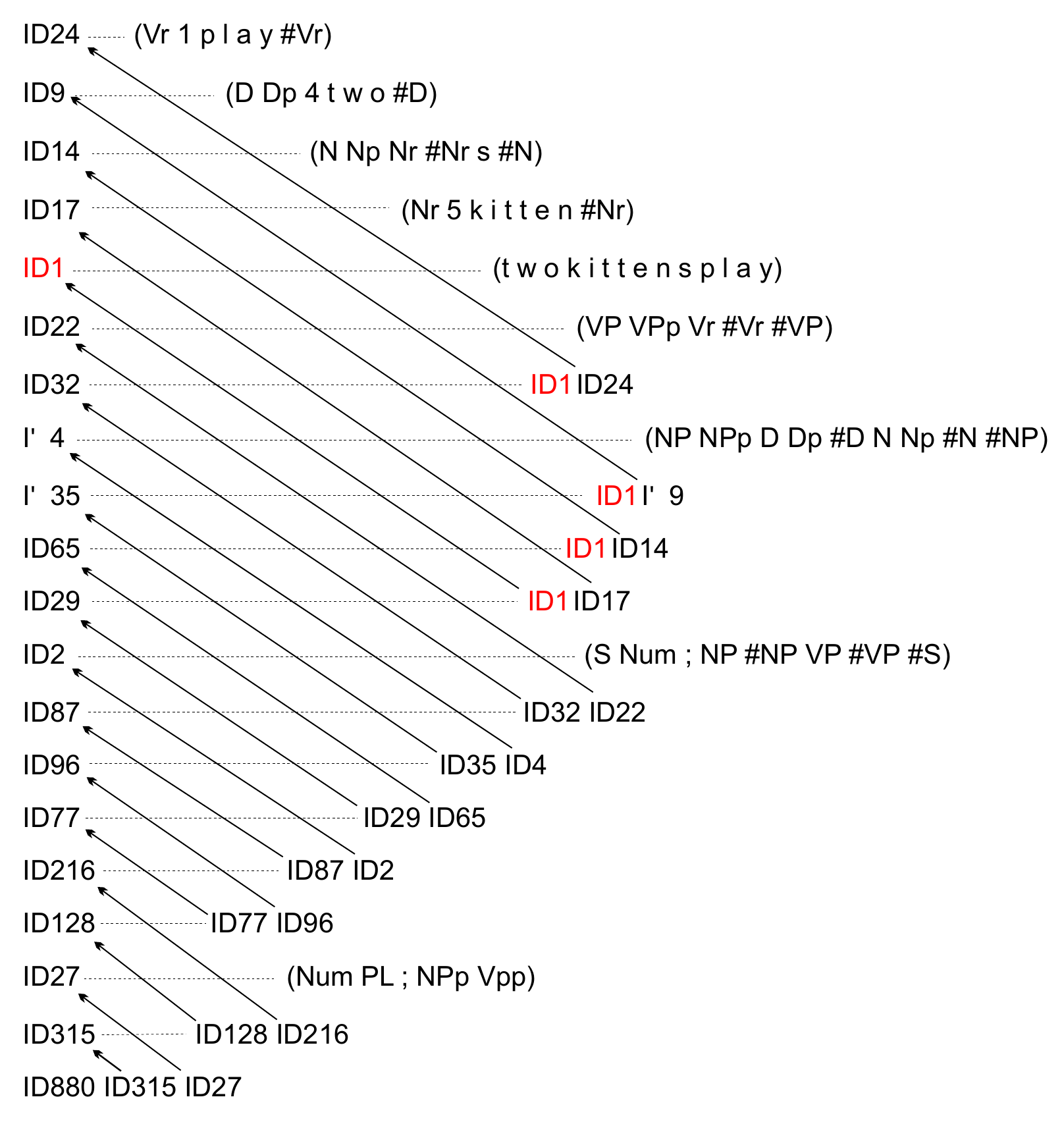}
\caption{An audit trail for the creation of the SPMA shown in Figure \ref{parsing-kittens-figure}. The line of text at the bottom of the figure shows, on the left, the identification number for the SPMA shown in Figure \ref{parsing-kittens-figure}. To the right of it are IDs of the two SPMA from which it was derived.
From each of the two IDs on middle and right of the first line, there is an arrow pointing to the same ID at the beginning of a line above. As before, this ID represents an SPMA, and to the right are two more IDs representing the two SPMAs from which it was derived. The remaining lines in the figure are the same except that, at higher levels, we begin to meet IDs that represent Old SP-patterns supplied to the SPCM at the beginning of processing. With the identifier `ID1', there are five instances in the figure (each one shown in red), so to avoid undue clutter, only one arrow is shown. Reproduced with permission from Figure 6 in \cite{trans-gran-2020}.}
\label{audit-trail-spma-figure}
\end{figure}

In summary:

\begin{itemize}

    \item All knowledge in the SPCM is represented by SP-patterns in structures that are familiar to people such as part-whole hierarchies, class-inclusion hierarchies, and more (see \cite[Section 5]{sp-micmup}).

    \item There is a very full audit trail for the creation of each SPMA. The structure of that audit trail is shown in Figure \ref{audit-trail-spma-figure}.

    \item There is also a very full audit trail from how the SPCM creates SP-grammars via unsupervised learning.

\end{itemize}

There is a fairly full discussion of these issues in \cite{trans-gran-2020}.

\subsection{Demonstration}

As an illustration of some of the transparency of the workings of the SP System, Figure \ref{sp-input-perspective-figure} shows how the SPMA in Figure \ref{parsing-kittens-figure} was created. The figure should be interpreted as described in its caption.

The SPMAs themselves are not shown in the figure but they are shown in the full audit trail from which the figure was derived, too big to be displayed here. That full audit trail contains much other information, including measures of IC associated with each SPMA, and absolute and conditional probabilities associated with each SPMA, calculated as described in \cite[Section 4.4]{sp-extended-overview} and \cite[Section 3.7]{wolff-2006}.

In a similar way, the process of unsupervised learning with the SPCM provides a very full audit trail for all intermediate structures that are created as learning proceeeds. As before this is much too big to be shown here.

\section{How to achieve probabilistic reasoning that integrates with other aspects of intelligence}\label{probabilistic-reasoning-section}

\begin{quote}

    ``What's going on now in the deep learning field is that people are building on top of these deep learning concepts and starting to try to solve {\em the classical AI problems of reasoning} and being able to understand, program, or plan.'' Yoshua Bengio \cite[p.~21]{ford-2018}, emphasis added.

\end{quote}

\begin{quote}

    ``A lot of people might [say]: `Deep learning systems are fine, but we don't know how to store knowledge, {\em or how to do reasoning}, or how to build more expressive kinds of models, because deep learning systems are just circuits, and circuits are not very expressive after all.'\thinspace'' Stuart J. Russell \cite[p.~49]{ford-2018}, emphasis added.

\end{quote}

\begin{quote}

    ``I think there's a presupposition that the way AIs can develop is by making individuals that are general-purpose robots like you see on Star Trek.~...~I ...~think, {\em in terms of general reasoning capacity, it's not going to happen for quite a long time}.'' Geoffrey Hinton \cite[p.~88]{ford-2018}, emphasis added.

\end{quote}

DNNs appear to be unsuitable for performing anything but the most rudimentary kind of reasoning. By contrast, a strength of the SP System is that, via the SPCM, several different kinds of probabilistic reasoning can be demonstrated, without any special provision or adaptation. (\cite[Section 10]{sp-extended-overview}, \cite[Chapter 7]{wolff-2006}).

The kinds of probabilistic reasoning that can be demonstrated with the SPCM are: one-step `deductive' reasoning; abductive reasoning, reasoning with probabilistic decision networks and decision trees, reasoning with `rules', nonmonotonic reasoning and reasoning with default values, causal diagnosis, and reasoning which is not supported by evidence. Also, the SPCM can function as an alternative to reasoning in Bayesian Networks, as, for example, in modelling the phenomenon of ``Explaining Away'', described by Pearl in \cite[pp.~7--9]{pearl-1997}. How that can be done is described in \cite[Section 10.2]{sp-extended-overview} and \cite[Section 7.8]{wolff-2006}.

Because of the probabilistic nature of the SP System (Appendix \ref{probabilistic-nature-of-sp-appendix}), all the kinds of reasoning that may be modelled in the SP System are fundamentally probabilistic, although it is possible to simulate the all-or-nothing nature of much classical logic when probabilities are at or near 0 or 1, as noted with references in Section \ref{symbolic-non-symbolic-section}.

As with the processing of natural language (Section \ref{integration-syntax-semantics-section}), a major strength of the SP System is that there can be smooth integration of diverse aspects of intelligence because all kinds of knowledge is represented with SP-patterns and all processing (apart from unsupervised learning) is achieved via the building of SPMAs.

This means that there can be smooth integration of one or more probabilistic kinds of reasoning with any or all of the SP System's strengths in several other aspects of AI and the representation of several kinds of knowledge.

\subsection{Demonstrations}

Examples of probabilistic reasoning by the SPCM are shown in (\cite[Section 10]{sp-extended-overview}, and there are more in \cite[Chapter 7]{wolff-2006}). These sources provide a fairly full picture of what can be done with the SPCM.

\begin{figure}[!hbt]
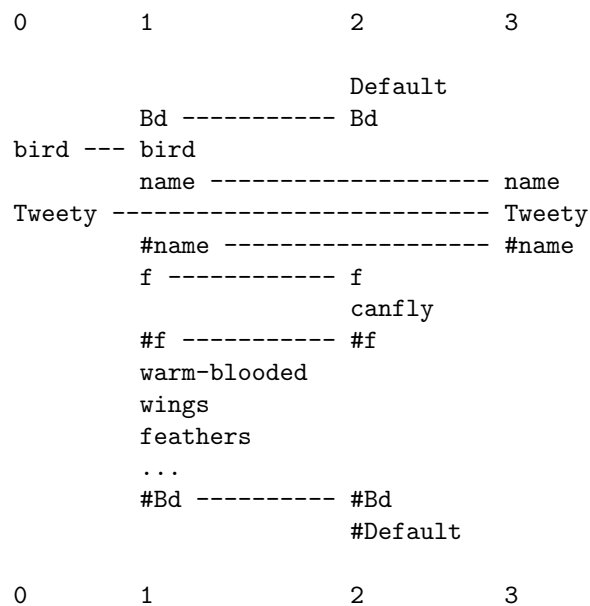

\fontsize{10.00pt}{12.00pt}
\centering
{\bf
\begin{BVerbatim}
0        1              2          3

                        Default
         Bd ----------- Bd
bird --- bird
         name -------------------- name
Tweety --------------------------- Tweety
         #name ------------------- #name
         f ------------ f
                        canfly
         #f ----------- #f
         warm-blooded
         wings
         feathers
         ...
         #Bd ---------- #Bd
                        #Default

0        1              2          3
\end{BVerbatim}
}
\caption{One of the three best SPMAs formed by the SPCM with the New SP-pattern `\texttt{bird Tweety}' and Old SP-patterns describing information about birds in general and more specific kinds of birds such as ostriches and penguins. Reproduced with permission from \cite[Figure 7.10]{wolff-2006}.}
\label{nonmon-figure-1}
\end{figure}

To give some of the flavour of how the SPCM may be applied to nonmonotonic reasoning, Figure \ref{nonmon-figure-1} shows one of the three best SPMAs created by the SPCM with the New SP-pattern `\texttt{bird Tweety}' (which appears in column 0, and which means that Tweety is a bird) and with a set of Old SP-patterns representing aspects of birds in general and also of specific kinds of birds such as ostriches and penguins.\footnote{This SPMA is arranged with SP-patterns in columns instead of rows, but otherwise it may be interpreted in exactly the same way as other SPMAs shown in this paper.}

From this SPMA and others formed at the same time, we may conclude that, with a relative probability of 0.66, Tweety can fly, but it is possible that Tweety, as an Ostrich ($p = 0.22$), or as a penguin ($p = 0.12$), would not be able to fly.

If the SPCM is run again with the New SP-pattern `\texttt{penguin Tweety}' (meaning that Tweety is a penguin), the best SPMA formed by the SPCM is shown in Figure \ref{nonmon-figure-2}. From this we may infer that Tweet would certainly not be able to fly ($p = 1.0$). Likewise when the New SP-pattern is `\texttt{ostrich Tweety}'.

\begin{figure}[!hbt]
\fontsize{10.00pt}{12.00pt}
\centering
\begin{BVerbatim}
0         1        2              3

                                  P
penguin ------------------------- penguin
                   Bd ----------- Bd
                   bird
          name --- name
Tweety -- Tweety
          #name -- #name
                   f ------------ f
                                  cannotfly
                   #f ----------- #f
                   warm-blooded
                   wings
                   feathers
                   ...
                   #Bd ---------- #Bd
                                  ...
                                  #P

0         1        2              3
\end{BVerbatim}
\caption{The best multiple alignment formed by SP61 with `penguin Tweety' in New and patterns in Old as described in the text. The relative probability of this multiple alignment is 1.0. Reproduced with permission from \cite[Figure 7.12]{wolff-2006}.}
\label{nonmon-figure-2}
\end{figure}

Thus, in accordance with the concept of nonmonotonic reasoning, the SPCM makes one set of inferences with the information that Tweety is a bird, but these inferences may be changed when we have more specific information.

\section{The need to re-balance research towards top-down strategies}\label{top-down-strategy-section}

\begin{quote}

    ``...~one of [the] stepping stones [towards progress in AI] would be an AI program that can really handle multiple, very different tasks. An AI program that's able to both do language and vision, it's able to play board games and cross the street, it's able to walk and chew gum. Yes, that is a joke, but I think it is important for AI to have the ability to do much more complex things.'' Oren Etzioni \cite[p.~502]{ford-2018}.

\end{quote}

\begin{quote}

    ``The central problem, in a word: current AI is {\em narrow}; it works for particular tasks that it is programmed for, provided that what it encounters isn't too different from what it has experienced before. That's fine for a board game like Go---the rules haven't changed in 2,500 years---but less promising in most real-world situations. Taking AI to the next level will require us to invent machines with substantially more flexibility. ...~To be sure, ...~narrow AI is certainly getting better by leaps and bounds, and undoubtedly there will be more breakthroughs in the years to come. But it's also telling: AI could and should be about so much more than getting your digital assistant to book a restaurant reservation.'' Marcus and Davis \cite[pp.~12--14]{marcus-davis-2019}, emphasis in the original.

\end{quote}

\begin{quote}

    ``Current approaches to machine learning lack important capabilities that will need to be developed to achieve artificial general intelligence. ...~creating new equations, new worldviews, new approaches to unfamiliar problems has so far been out of reach for computers. ...~we do not have a reasonably complete theory of general intelligence in people, let alone in machines. ...~Today's approaches to artificial intelligence have been extremely successful at creating hedgehogs, but general intelligence requires foxes.'' Herbert Roitblat \cite[p.~277]{roitblat-2020}.

\end{quote}

The first quote is, in effect, a call for a top-down strategy in AI research, developing a theory or theories that can be applied to a range of phenomena, not just one or two things in a narrow area. The potential advantages of that kind of strategy in terms of the generality of theories, and their value in terms of Ockham's razor, are described in Section  \ref{advantages-of-top-down-strategy-section}, below.

In this connection, the SP System performs well. Appendix \ref{simplicity-power-appendix} describes how {\em the SP System has adopted a unique top-down strategy: attempting SIABC}. And it seems fair to say that development of the SP System, with the powerful concept of SPMA at its core, has achieved a favourable combination of conceptual Simplicity with descriptive and explanatory Power ({\em ibid.}).

\subsection{Some advantages of a top-down strategy}\label{advantages-of-top-down-strategy-section}

Here are some key features of a top-down strategy in research, and their potential benefits:

\begin{enumerate}

    \item {\em Broad reach}. As noted above, achieving generality requires that the data from which a theory is derived should have a broad reach, like the overarching goal of the SP programme of research, SIABC, summarised at the beginning of Appendix \ref{simplicity-power-appendix}.

     \item {\em Ockham's razor, Simplicity and Power}. That broad scope is important for two reasons:

        \begin{itemize}

            \item In accordance with Ockham's razor, a theory should be as {\em Simple} as possible but, at the same time, it should retain as much as possible of the descriptive and explanatory {\em Power} of the data from which it is derived (Appendix \ref{simplicity-power-appendix}).

            \item But those two measures are more important when they apply to a wide range of phenomena than when they apply only to a small piece of data.

        \end{itemize}

        Of course, a top-down strategy does not guarantee that one will achieve a favourable combination of Simplicity with Power. But it is a useful first step.

    \item {\em If you can't solve a problem, enlarge it}. A broad scope, as above, can be challenging, but it can also make things easier. Thus President Eisenhower is reputed to have said: ``If you can't solve a problem, enlarge it'', meaning that putting a problem in a broader context may make it easier to solve. Good solutions to a problem may be hard to see when the problem is viewed through a keyhole, but become visible when the door is opened.

    \item {\em Micro-theories rarely generalise well}. Apart from the potential value of `enlarging' a problem (point 2 above), and broad scope (point 1), a danger of adopting a narrow scope is that any micro-theory or theories that one may develop for that narrow area are unlikely to generalise well to a wider context---with correspondingly poor results in terms of Simplicity and Power.

        For reasons of that kind, Allen Newell, in his famous essay ``You can't play 20 questions with nature and win'' \cite{newell-1973}, urges researchers in psychology to develop theories with wide scope (pp.~284--289), and, accordingly, to work with ``a genuine slab of human behaviour'' (p.~303). This kind of thinking is the basis of his book {\em Unified Theories of Cognition} \cite{newell-1990}.

    \item {\em Bottom-up strategies and the fragmentation of research}. The prevailing view about how to reach AGI seems to be ``...~that we'll get to general intelligence step by step by solving one problem at a time.'' expressed by Ray Kurzweil \cite[p.~234]{ford-2018}. And much research in AI has been, and to a large extent still is, working within this kind of bottom-up strategy: developing ideas in one area, and then trying to generalise them to another area, and so on.

        But it seems that in practice the research rarely gets beyond two areas, and, as a consequence, there is much fragmentation of research (next).

\end{enumerate}

In connection with the fragmentation of research (point 5, above), John Kelly and Steve Hamm (both of IBM) write:

\begin{quote}

    ``Today, as scientists labor to create machine technologies to augment our senses, there's a strong tendency to view each sensory field in isolation as specialists focus only on a single sensory capability. Experts in each sense don't read journals devoted to the others senses, and they don't attend one another's conferences. Even within IBM, our specialists in different sensing technologies don't interact much.'' \cite[location 1004]{kelly-hamm-2013}.

\end{quote}

\noindent And Pamela McCorduck writes:

\begin{quote}

    ``The goals once articulated with debonair intellectual verve by AI pioneers appeared unreachable ...~Subfields broke off---vision, robotics, natural language processing, machine learning, decision theory---to pursue singular goals in solitary splendor, without reference to other kinds of intelligent behaviour.'' \cite[p.~417]{mccorduck-2004}. Later, she writes of ``the rough shattering of AI into subfields ...~and these with their own sub-subfields---that would hardly have anything to say to each other for years to come.'' \cite[p.~424]{mccorduck-2004}. She adds: ``Worse, for a variety of reasons, not all of them scientific, each subfield soon began settling for smaller, more modest, and measurable advances, while the grand vision held by AI's founding fathers, a general machine intelligence, seemed to contract into a negligible, probably impossible dream.'' ({\em ibid.}).

\end{quote}

Attempts to overcome these problems have given rise to two main strands of work: 1) research mentioned above inspired by Newell's {\em Unified Theories of Cognition} \cite{newell-1990}; and 2) research aiming to develop AGI (see, for example, \cite{ikle-etal-2018}). But, while both strands of research are welcome, it seems that neither of them have yet managed to escape properly from the problems of bottom-up research.

That a top-down approach to the development of a fully-integrated AGI is proving difficult is suggested by observations quoted in Section \ref{fresh-approach-simplification-integration-section}, below. Attempts to develop unified theories of cognition, or AGI, have, so far, not overcome the fragmentation of AI that is so well described by Pamela McCorduck \cite[p.~417]{mccorduck-2004}, as quoted above. And it seems that part of the problem is the reluctance of researchers to break free from a bottom-up strategy in research.


\subsection{A fresh approach}\label{fresh-approach-section}

Writing about possible ways forward, Marcus and Davis say:

\begin{quote}

    ``What's missing from AI today---and likely to stay missing, until and unless the field takes a fresh approach---is {\em broad} (or `general') intelligence. AI needs to be able to deal not only with specific situations for which there is an enormous amount of cheaply obtained relevant data, but also problems that are novel, and variations that have not been seen before.

    ``Broad intelligence, where progress has been much slower, is about being able to adapt flexibly to a world that is fundamentally open-ended---which is the one thing humans have, in spades, that machines haven't yet touched. But that's where the field needs to go, if we are to take AI to the next level.'' (\cite[p.~15]{marcus-davis-2019}, emphasis in the original).

\end{quote}

\noindent And:

\begin{quote}

    ``We call this book {\em Rebooting AI} because we believe that the current approach isn't on a path to get us to AI that is safe, smart, or reliable. A short-term obsession with narrow AI and the easily achievable `low-hanging fruit' of big data has distracted too much attention away from a longer-term and much more challenging problem that AI needs to solve if it is to progress: the problem of how to endow machines with a deeper understanding of the world. Without that deeper understanding, we will never get to truly trustworthy AI. In the technical lingo, we may be stuck at a local maximum, an approach that is better than anything similar that's been tried, but nowhere good enough to get us where we want to go.

    ``For now, there is an enormous gap---we call it `the AI Chasm'---between ambition and reality.'' (\cite[pp.~17--18]{marcus-davis-2019}).

\end{quote}

Of course, the SP System, like other AI systems, does not immediately deliver `broad AI' as described by Marcus and Davis. But, for reasons summarised in the following subsections, it may with some justice claim to be a promising foundation for the development of AGI, and it may claim to be the kind of ``fresh approach'', with the generality that they call for.

\subsubsection{A top-down strategy for the development of the SP System}

As noted in Section \ref{top-down-strategy-section}, the SP programme of research is one of very few to have adopted a top-down strategy, and it seems to have achieved more in that paradigm than any other.

As noted in Appendix \ref{simplicity-power-appendix}, the overarching goal of the SP programme of research is simplification and integration of observations and concepts across AI, mainstream computing, mathematics, and human learning, perception and cognition (SIABC).

\subsubsection{Aiming for a theory that combines conceptual {\em Simplicity} with descriptive and explanatory {\em Power}}\label{fresh-approach-simplification-integration-section}

It seems that the simplicity-with-power objective mentioned in the previous subsection is largely met by the SP System, as described here.

{\em Simplicity}. The SPCM comprises mainly the processes for SPMA (Appendix \ref{sp-multiple-alignment-appendix}), with processes for unsupervised learning (Appendix \ref{sp-unsupervised-learning-appendix}). Overall, the model is remarkably simple, considering its versatility (next).

{\em Power}. The descriptive and explanatory Power of the SP System lies in its versatility with aspects of intelligence (Appendix \ref{versatility-in-aspects of-intelligence-appendix}), including diverse forms of reasoning (Appendix \ref{versatility-in-reasoning-appendix}), its versatility in the representation of diverse kinds of knowledge (Appendix \ref{versatility-in-representation-of-knowledge-appendix}), and its facility for the seamless integration of diverse aspects of intelligence and diverse kinds of knowledge, in any combination (Appendix \ref{seamless-integration-appendix}).

{\em Related research}. It is true that there has been research for many years inspired by Allen Newell's book about {\em Unified Theories of Cognition} \cite{newell-1990}, and there is also a significant strand of research aiming to develop `Artificial General Intelligence' (see, for example, \cite{ikle-etal-2018}).

But despite the welcome aims of researchers in these areas, it appears fair to say that broad AI or AGI has not yet been achieved as described in these quotes:

\begin{quote}

    ``Despite all the current enthusiasm in AI, the technologies involved still represent no more than advanced versions of classic statistics and machine learning. Behind the scenes, however, many breakthroughs are happening on multiple fronts: in unsupervised language and grammar learning, deep-learning, generative adversarial methods, vision systems, reinforcement learning, transfer learning, probabilistic programming, blockchain integration, causal networks, and many more.'' \cite[Preface, Location 51]{ikle-etal-2018}.

\end{quote}

\noindent And:

\begin{quote}

    ``Computers have not yet achieved artificial general intelligence not because they lack some ineffable property that humans have, such as consciousness, but because computer scientists have not been designing for general intelligence.'' \cite[p.~314]{roitblat-2020}.

\end{quote}

\begin{quote}

    ``If we want humanlike intelligence, we must figure out a way to construct it from the tools that we have available or we must build new tools. There are some attempts to create general intelligence with current tools, but none of them, so far, has demonstrated any success.'' \cite[p.~328]{roitblat-2020}.

\end{quote}

Since, as noted above, the SP System has already largely met the objective of combining Simplicity with Power (Section \ref{fresh-approach-simplification-integration-section}), it seems fair to say that the SP System may claim to be part of a fresh approach to the development of broad AI.

\subsubsection{Information compression as a unifying principle in the SP System}\label{fresh-approach-ic-as-unifying-principle-section}

Other reasons to be optimistic about the potential of the SP System to provide a relatively firm foundation for the development of broad AI are:

\begin{itemize}

    \item That all processing in the SP System is achieved via IC;

    \item That IC lies at the heart of how the system represents its knowledge;

    \item That IC is an extremely general concept that can in principle be applied to any kind of processing or the representation of any kind of knowledge;

    \item That evidence summarised in Appendix \ref{sp-str-pot-appendix} shows the wide range of kinds of processing and kinds knowledge where IC may be applied to good effect.

    \item To be more specific, IC in the SP System is achieved largely via SPMA (Appendix \ref{sp-multiple-alignment-appendix}), and via unsupervised learning (Appendix \ref{sp-unsupervised-learning-appendix}). These two processes exemplify the generality of IC in the workings of the SP System.

    \item And perhaps most significantly, there is much evidence for the importance of IC in the workings of brains and nervous systems \cite{sp-compression}.

\end{itemize}

\subsubsection{Generalisation
 and the correction of over- and under-generalisations}\label{fresh-approach-generalisation-section}

A potentially valuable bonus from a theory of HLPC and AI which is founded on IC, for both the representation and the processing of knowledge, is that it provides a robust, coherent theory of generalisation and the correction of both over- and under-generalisations, which has some empirical support (Section \ref{generalisation-section}).

As described in Section \ref{driverless-cars-section}, such a theory of generalisation may helpg to overcome problems of adaptability in driverless cars.

The variety of situations that a driver may encounter is far too large for any driver (human or computer) to achieve competence by learning specific responses to specific situations. In much the same way that we learn our native language or languages, learner drivers must generalise beyond their experience to date, but they must not overgeneralise. And it seems that IC provides the key.

\subsubsection{Potential benefits and applications of the SP System}\label{fresh-approach-apps-section}

Another means of assessing the generality of the SP System is via potential benefits and applications of the system which are credible in the sense that they are not mere speculations but have supporting evidence via the workings and performance of the system.

In this respect, the SP System scores well, as can be seen from the range of potential benefits and applications described in \cite{sp-benefits-apps} and in several other papers that may be downloaded via links from \href{http://www.cognitionresearch.org/sp.htm}{cognitionresearch.org/sp.htm}, including peer-reviewed published papers on the application of the SP System to problems in: big data \cite{sp-big-data}, the development of human-like intelligence in autonomous robots \cite{sp-autonomous-robots}, computer vision and natural vision \cite{sp-vision}, intelligent databases \cite{sp-intelligent-database}, medical diagnosis \cite{sp-medical-diagnosis}, and mathematics \cite{sp-micmup}.

\subsection{Foxes not hedgehogs}

Although we may quibble with Roitblat's suggestion that computers of today cannot create new equations, he is certainly right that there is no widely-agreed definition of general intelligence.

In the SP programme of research, the focus has been on creating capabilities that are agreed to be part of general intelligence---unsupervised learning, the processing of natural language, pattern recognition, several kinds of reasoning, and more, summarised in Appendix \ref{sp-str-pot-appendix}.

For a system that is essentially quite simple, its versatility is good. But there are other challenges in, for example, how to model aspects of commonsense knowledge and commonsense reasoning (Section \ref{csrk-section}).

\section{How to minimise the risk of accidents with self-driving vehicles}\label{driverless-cars-section}

\begin{quote}

    ``In the early versions of Google's [driverless] car, ...~the problem was that every day, Google found themselves adding new rules. Perhaps they would go into a traffic circle ...~and there would be a little girl riding her bicycle the wrong way around the traffic circle. They didn't have a rule for that circumstance. So, then they have to add a new one, and so on, and so on.'' Stuart J. Russell \cite[p.~47]{ford-2018}.

\end{quote}

\begin{quote}

    ``...~the principal reason [for pessimism about the early introduction of driverless cars for all situations is] that if you're talking about driving in a very heavy metropolitan location like Manhattan or Mumbai, then the AI will face a lot of unpredictability. It's one thing to have a driverless car in Phoenix, where the weather is good and the population is a lot less densely packed. The problem in Manhattan is that anything goes at any moment, nobody is particularly well-behaved and everybody is aggressive, the chance of having unpredictable things occur is much higher.'' Gary Marcus \cite[p.~321]{ford-2018}.

\end{quote}

A na{\"i}ve approach to the avoidance of accidents with driverless cars would be to specify stimulus-response pairs, where the stimulus would be a picture of the road in front (perhaps including sounds), and the response would be a set of actions with the steering wheel, brakes, and so on. Of course, driving is far too complex for anything like that to be adequate.

It seems that, for any kind of driver, either human or artificial, some kind of generalisation from experience is essential (Section \ref{generalisation-section}). In that connection, people will have the benefit of all their visual experience prior to their driving lessons, but the same principles apply.

If a person or a driverless car has learned to apply the brakes when a child runs out in front, that learning should be indifferent to the multitude of images that may be seen: the child may be fat or thin; tall or short; running, skipping, or jumping; in a skirt or wearing trousers; facing towards the car or away from it; seen through rain or not; lit by street lights or by the sun; and so on.

There may be some assistance from `generalisation via perception' (Section \ref{generalisation-via-perception-section}) but that in itself is unlikely to be sufficient. It seems that something like `generalisation via unsupervised learning' (Section \ref{generalisation-via-unsupervised-learning-section}) is also needed.

With those two kinds of generalisation, it seems possible that, with reasonable amounts of driving experience across a range of driving conditions, the risk of accidents may be minimised.

As with human drivers, there would still be errors made by the artificial driver---because the generalisations would be probabilistic. But there is potential for the artificial driver to do substantially better than most human drivers---by inheriting the experience of many other artificial drivers, by not suffering from such things as falling asleep at the wheel, and by not being tempted to consume alcohol before driving.

\section{The need for strong compositionality in the structure of knowledge}\label{strong-compositionality-section}

\begin{quote}

    ``By the end of the '90s and through the early 2000s, neural networks were not trendy, and very few groups were involved with them. I had a strong intuition that by throwing out neural networks, we were throwing out something really important.

    ``Part of that was because of something that we now call compositionality: The ability of these systems to represent very rich information about the data in a compositional way, where you compose many building blocks that correspond to the neurons and the layers.'' Yoshua Bengio \cite[p.~25]{ford-2018}.

\end{quote}

The neurons and layers of a DNN may be seen as building blocks for a concept, and may thus be seen as an example of compositionality. But it seems that any such view of the layers in a DNN is weak, with many exceptions to any strict compositionality. In general, DNNs fail to capture the way in which we conceptualise a complex thing like a car in terms of smaller things (engine, wheels, etc), and these in terms of still smaller things (pistons, valves, etc), and so on. This kind of hierarchical representation of concepts, which is prominent in the way people conceptualise things, we may call `strong' compositionality.

It appears that in this connection, the SP System has a striking advantage compared with DNNs. Any SP-pattern may contain SP-symbols that serve as references to other SP-patterns, a mechanism which allows part-whole hierarchies and class-inclusion hierarchies to be built up through as many levels as are required ((\cite[Section 9.1]{sp-extended-overview}, \cite[Section 6.4]{wolff-2006})).

\subsection{Demonstrations}

This kind of compositionality can be seen in Figure \ref{parsing-kittens-figure}, where the SP-pattern `\texttt{D Dp 4 t w o \#D}' (which represents a word), and this connects with the SP-pattern `\texttt{NP NPp D Dp \#D N Np \#N \#NP}' which which represents a noun phrase, and that connects with the SP-pattern `\texttt{VP VPp Vr \#Vr \#VP}' which represents a verb phrase, and that connects with the SP-pattern `\texttt{S Num ; NP \#NP VP \#VP \#S}' which represents a sentence.

Apart from part-whole hierarchies, the SP System also lends itself to the representation and processing of class-inclusion hierarchies, as can be seen in \cite[Figure 16, Section 9.1]{sp-extended-overview}.

The `DONSVIC' principle, mentioned in Section \ref{symbolic-non-symbolic-section}, which is the `Discovery of Natural Structures via Information Compression') \cite[Section 5.2]{sp-extended-overview}, may provide further evidence for composionality in the SPCM when it is more fully developed.

It is envisaged that, with appropriate data, the SPCM will demonstrate the discovery of words and similar chunks of information (as it does now) and also the grouping of such structures into part-whole hierarchies or class-inclusion hierarchies---in accordance with the DONSVIC principle.

\section{The challenges of commonsense reasoning and commonsense knowledge}\label{csrk-section}

\begin{quote}

    ``We don't know how to build machines that have human-like common sense. We can build machines that can have knowledge and information within domains, but we don't know how to do the kind of common sense we all take for granted.'' Cynthia Breazeal \cite[p.~456]{ford-2018}.

\end{quote}

\begin{quote}

    ``We still don't have any real AI in the sense of the original vision of the founders of the field, of what I think you might refer to as AGI—machines that have that same kind of flexible, general-purpose, common sense intelligence that every human uses to solve problems for themselves.'' Joshua Tenenbaum \cite[p.~472]{ford-2018}.

\end{quote}

\begin{quote}

    ``To achieve human-level performance in domains such as natural language processing, vision, and robotics, basic knowledge of the commonsense world---time, space, physical interactions, people, and so on---will be necessary.

    ``Although a few forms of commonsense reasoning, such as taxonomic reasoning and temporal reasoning are well understood, progress has been slow.'' \cite[Key insights]{davis-marcus-2015}.

\end{quote}

Although `commonsense reasoning' (CSR) is a kind of reasoning, it is discussed here, with `commonsense knowledge' (CSK), in a section that is separate from Section \ref{probabilistic-reasoning-section} (about the strengths and potential of the SPCM for several kinds of probabilistic reasoning). This is because of the way CSR and CSK (which, together, may be referred to as `CSRK') have been developing as a discrete subfield of AI (see, for example, \cite{davis-marcus-2015}).

Judging by the nature of DNNs and the paucity of research on how they might be applied in CSRK research \cite{schmidhuber-2015}, it seems that DNNs are not well suited to this aspect of AI. But preliminary studies suggest that the SP System has potential in this area.

The SP System may prove useful with CSRK \cite[Section 3]{spcsrk2-2019}, and more so when `unfinished business' in the development of the SPCM has been completed (Appendix \ref{unfinished-business-appendix}).

\subsection{Demonstrations}

Aspects of CSRK may be modelled with the SPCM \cite[Sections 4 to 6]{spcsrk2-2019}: how to interpret a noun phrase like ``water bird''; how, under various scenarios, to assess the strength of evidence that a given person committed a murder; how to interpret the horse's head scene in {\em The Godfather} film.

A fourth problem---how to model the process of cracking an egg into a bowl---is beyond what can be done with the SP System as it is now \cite[Section 9]{spcsrk2-2019}, but fixing the problems mentioned Appendix \ref{unfinished-business-appendix} may make it feasible.

With the SPCM, it is possible to determine the referent of an ambiguous pronoun in a `Winograd schema' type of sentence \cite{sp-winograd-schemas}, where a Winograd schema is a pair of sentences like {\em The city councilmen refused the demonstrators a permit because they feared violence} and {\em The city councilmen refused the demonstrators a permit because they advocated revolution}, and the ambiguous pronoun in each sentence is ``they'' \cite{levesque-2011}.

\section{Establishing the key importance of information compression in AI research}\label{central-role-for-ic-section}

There is little about information compression in {\em Architects of Intelligence}, except for some brief remarks about autoencoders by Yoshua Bengio in \cite[p.~26]{ford-2018}:

\begin{quote}

    ``Autoencoders have changed quite a bit since [the] original vision. Now, we think of them in terms of taking raw information, like an image, and transforming it into a more abstract space where the important, semantic aspect of it will be easier to read. That's the encoder part. The decoder works backwards, taking those high-level quantities—that you don't have to define by hand—and transforming them into an image. That was the early deep learning work.

\end{quote}

Now it seems that interest in autoencoders has waned: ``...~a few years later, we discovered that we didn't need these approaches to train deep networks, we could just change the nonlinearity.'' Yoshua Bengio \cite[p.~26]{ford-2018}.

This fairly relaxed view of IC in AI research, as described by a leading researcher in AI, and the rather low profile of IC in a wide-ranging review of research on DNNs \cite{schmidhuber-2015}, contrasts with the central role in the SP programme of research:

\begin{enumerate}

    \item IC is fundamental:

        \begin{itemize}

            \item In HLPC \cite{sp-compression};

            \item In mathematics \cite{sp-micmup};

            \item In the design of the SP System \cite{sp-extended-overview,wolff-2006};

        \end{itemize}

    \item With regard to the prominence of IC in HLPC, one of the more persuasive reasons for recognising this feature of brains and nervous systems is that, in many creatures, IC is likely to have been a product of natural selection \cite[Section 4]{sp-compression}.

    \item IC provides several reasons to be optimistic about the potential of the SP System to provide a relatively firm foundation for the development of human-like general AI (Section \ref{fresh-approach-ic-as-unifying-principle-section}).

\end{enumerate}

With regard to the second point, Marcus provides persuasive evidence in his book {\em Kluge} \cite{marcus-2008}, that the haphazard nature of natural selection produces kluges in human thinking, meaning clumsy makeshift solutions that nevertheless work. But, at the same time, there is compelling evidence for the importance of IC in the workings of brains and nervous systems \cite{sp-compression}. Probably, the two ideas are both true.

\section{Establishing the importance of a biological perspective in AI research}\label{biological-perspective-section}

Amongst the researchers interviewed by Ford, and by Ford himself, there is little or no acknowledgement of the importance of a biological perspective in AI. Some remarks in this area include:

\begin{quote}

    ``Artificial neural networks, in which software roughly emulates the structure and interaction of biological neurons in the brain, date back at least to the 1950s. Simple versions of these networks are able to perform rudimentary pattern recognition tasks, and in the early days generated significant enthusiasm among researchers. ...'' Martin Ford \cite[location 84]{ford-2018}.

\end{quote}

\begin{quote}

    ``Deep learning will do some things, but biological systems rely on hundreds of algorithms, not just one algorithm. [AI researchers] will need hundreds more algorithms before we can make that progress, and we cannot predict when they will pop.'' Rodney Brooks \cite[p.~427]{ford-2018}.

\end{quote}

In the first quote, the main idea is that artificial neural networks are only rough imitations of real neurons, with nothing to suggest whether or how better models might be created.

What Rodney Brooks describes in the second quote is much like Marvin Minsky's concept of diverse agents \cite{minsky-1986} as the basis for AI. It seems that both of them are unfalsifiable theories because, for every attempt to prove the theory wrong, a new algorithm may be added to plug the gap. And that is likely to mean a theory with ever-decreasing merit in terms of Ockham's razor. Nevertheless, a biological perspective can yield important insights for AI and how it may be developed (Section \ref{central-role-for-ic-section}).

By contrast with the `many algorithms' perspectives of Brooks and Minsky, the SP programme of research has, from the beginning, been tightly focussed on the importance of IC in the workings of brains and nervous systems \cite{sp-compression} and the corresponding importance of IC as a unifying principle in AI (Section \ref{central-role-for-ic-section}). Much of the evidence in \cite{sp-compression} comes from neuroscience, psychology, and other aspects of biology.

\section{Establishing whether knowledge in the brain is represented in `distributed' or `localist' form}\label{distributed-localist-representations-section}

\begin{quote}

    ``In a hologram, information about the scene is distributed across the whole hologram, which is very different from what we're used to. It's very different from a photograph, where if you cut out a piece of a photograph you lose the information about what was in that piece of the photograph, it doesn't just make the whole photograph go fuzzier.'' Geoffrey Hinton \cite[p.~79]{ford-2018}.

\end{quote}

A persistent issue in AI and theories of HLPC is whether knowledge in the brain is represented in a `distributed' or `localist' form.

In DNNs, knowledge is distributed in the sense that it is encoded in the strengths of connections between many neurons across several layers of each DNN. Since DNNs provide the most fully developed examples of AI systems with distributed knowledge, it is assumed in present discussion that they are representative for that kind of artificial system for representing knowledge.

In SP-Neural, a `neural' version of the SP System (Appendix \ref{sp-neural-appendix}), knowledge is localised in the sense that there may be a single neuron, or more likely a small cluster of neurons, representing any one concept such as `my house', but such a concept is likely to have links to many other concepts in other places such as `roof', `window', `doors', and so on.

This issue is essentially the much-debated issue of whether the concept of `my grandmother' is represented in one place in one's brain or whether the concept may be represented via a diffuse collection of neurons throughout the brain.

Although the following conclusion will not be free from controversy, it seems that the weight of evidence now favours a localist view, and the SP System is in keeping with that conclusion:

\begin{itemize}

    \item Mike Page \cite[pp.~461--463]{page-2000} discusses several studies that provide direct or indirect evidence in support of localist encoding of knowledge in the brain.

    \item The SP System, in both its abstract form (Appendix \ref{outline-of-sp-system-appendix}) and as SP-Neural (Appendix \ref{sp-neural-appendix}), is unambiguously localist. To the extent that the SP System provides a plausible framework for the development of AGI, it provides evidence in support of localist forms for knowledge.

    \item Since DNNs are vulnerable to the problem of catastrophic forgetting (Section \ref{catastrophic-forgetting-section}), this seems to be a problem more generally for distributed representations of knowledge. With knowledge stored in the strengths of connections between neurons, there must be a risk that two or more concepts will interfere with each other.

    \item It is true that if knowledge of one's grandmother is contained within a single neuron, death of that neuron would destroy one's knowledge of one's grandmother. But:

        \begin{itemize}

            \item As Barlow points out \cite[pp. 389--390]{barlow-1972}, a small amount of replication will give considerable protection against this kind of catastrophe.

            \item Any person who has suffered a stroke, or is suffering from dementia, may indeed lose the ability to recognise close relatives or friends.

        \end{itemize}

    \item Is it conceivable that, with a localist representation, there are enough neurons in the human brain to store the knowledge that a typical person, or, more to the point, an exceptionally knowledgeable person, may have? Arguments and calculations relating to this issue suggest that it is indeed possible for us to store what we know in localist form, and with substantial room to spare for multiple copies \cite[Section 11.4.9]{wolff-2006}. A summary of the arguments and calculations is in \cite[Section 4.4]{spneural-2016}.

    \item In support of the conclusion in the preceding point, the key importance of IC in the organisation and workings of the SP System (Appendix \ref{two-mechanisms-for-ic-appendix}) provides a good reason for supposing that, in a mature SP Machine\footnote{Developed, perhaps, via the programme of research described in \cite{sp-palade-wolff}}, knowledge acquired via unsupervised learning will be stored in a highly compressed form.

\end{itemize}

\section{How to bypass the limited scope for adaptation in deep neural networks}\label{scope-for-adaptation-section}

This and the following two sections describe problems in AI that are not apparently considered in \cite{ford-2018}, except briefly, but are significant problems in AI research that the SP System may solve.

The problem considered here is that, contrary to how DNNs are normally viewed, they are relatively restricted in their scope for adaptation.

\subsection{Any one DNN is designed to learn a single concept}

Although each DNN is designed to learn one concept, one could provide multiple DNNs for the learning of multiple concepts. But, since a DNN has multiple layers and multiple connections between layers (which is what makes it `deep'), the provision of a DNN for each of the many concepts that people can learn would be expensive in terms of neural structures and neural functioning.

In the SPCM, the concept of SP-pattern (with the concept of SPMA), provides much greater scope for modelling the world than the relatively constrained framework of DNNs. This is because each concept is represented by one SP-pattern, there is no limit to the number of SP-patterns that may be formed (apart from the memory that is available in the host computer), and there is no limit to the number of ways in which a given SP-pattern may be connected to other SP-patterns within the framework of SPMAs (in much the same way that there is no limit to the number of ways in which a given web page may be connected to other web pages).

\subsection{The constraints of layers and connections}

The layers in a DNN, and the potential connections amongst them, are finite and pre-defined \cite[Section 1]{schmidhuber-2015}.

By contrast, there is no limit to the variety of concepts that may be encoded in SP-patterns. And there is no limit to the variety of connections that may exist between SP-patterns.

\section{How to eliminate the problem of catastrophic forgetting}\label{catastrophic-forgetting-section}

\begin{quote}

    ``We find that the CF [catastrophic forgetting] effect occurs universally, without exception, for deep LSTM-based [Long Short-Term Memory based] sequence classifiers, regardless of the construction and provenance of sequences. This leads us to conclude that LSTMs, just like DNNs [Deep Neural Networks], are fully affected by CF, and that further research work needs to be conducted in order to determine how to avoid this effect (which is not a goal of this study).'' Monika Schak and Alexander Gepperth \cite{schak-gepperth-2019}.

\end{quote}

Catastrophic forgetting---which is a problem for at least some DNNs---is the way in which, when a given DNN has learned one thing and then it learns something else, the new learning wipes out the earlier learning (see, for example, \cite{goodfellow-etal-2015}). This problem, mentioned at the end of the section about transfer learning (Section \ref{transfer-learning-section}), is quite different from human learning, where new learning normally builds on earlier learning, as described in Section \ref{transfer-learning-section}, although of course we all have a tendency to forget some things.

The SP System is entirely free of the problem of catastrophic forgetting. The reasons that, in general, DNNs suffer from catastrophic forgetting and that the SP System does not, are that:

\begin{itemize}

    \item In DNNs there is a single structure for the learning and storage of new knowledge, a concept like `my house' is encoded in the strengths of connections between artificial neurons in that single structure, so that the later learning of a concept like `my car' is likely to disturb the strengths of connections for `my house' (see the discussion of `grandmother' cells in \cite[Section 5.8]{spneural-2016});

        An apparent way round this problem is to provide an extremely large network for training, with multiple layers in all parts of the network, so that very large patterns may be used in training. There are implications here for the computational resources that are required (Section \ref{comp-resources-speed-volumes-of-data-section}).

    \item By contrast, the SP System has an SP-pattern for each concept in its repository of knowledge, there is no limit to the number of such SP-patterns that may be stored (apart from the limit imposed by the available storage space in the computer and associated information storage), and, although there may be many connections between SP-patterns, there is no interference between any one SP-pattern and any other.

\end{itemize}

It is true that, in the SP System, a concept like `person' may be composed of subsidiary concepts like `head', `body' and `legs' and that corruption of those subsidiary concepts would corrupt the concept of `person'. But, in accordance with our ordinary experience, it is entirely feasible to provide an SP-pattern to record that `John' has an injury to his `head' without disturbing the SP-pattern that records the fact that the `head' of a typical `person' is not injured.

\section{Motivations and emotions}\label{motivations-emotions-section}

\begin{quote}

    ``How much prior structure do we need to build into those systems for them to actually work appropriately and be stable, and for them to have intrinsic motivations so that they behave properly around humans? There's a whole lot of problems that will absolutely pop up, so AGI might take 50 years, it might take 100 years, I'm not too sure.'' Yann LeCun \cite[p.~130]{ford-2018}.

\end{quote}

\begin{quote}

    ``Machine learning needs a lot of data, and so I borrowed [a] dataset [from Cambridge Autism Research Center] to train the algorithms I was creating, on how to read different emotions, something that showed some really promising results. This data opened up an opportunity to focus not just on the happy/sad emotions, but also on the many nuanced emotions that we see in everyday life, such as confusion, interest, anxiety or boredom.'' Rana el Kaliouby \cite[p.~209]{ford-2018}.

\end{quote}

\begin{quote}

    ``[A] subtle question is that of relating emotionally to other beings. I'm not sure that's even well defined, because as a human you can fake it. There are people who fake an emotional connection to others. So, the question is, if you can get a computer to fake it well enough, how do you know that's not real?'' Daphne Koller \cite[p.~394]{ford-2018}.

\end{quote}

\begin{quote}

    ``If you look at human intelligence we have all these different kinds of intelligences, and social and emotional intelligence are a profoundly important, and of course underlies how we collaborate and how we live in social groups and how we coexist, empathize, and harmonize.'' Cynthia Breazeal \cite[p.~450]{ford-2018}.

\end{quote}

\begin{quote}

    ``...~why are we assuming the same evolutionary forces that drove the creation of our motivations and drives would be anything like those of [a] super intelligence?'' Cynthia Breazeal \cite[p.~457]{ford-2018}.

\end{quote}

In developing AGI, motivations and emotions are clearly important, not least because of the possibility that super-intelligent AIs might come to regard people as dispensable. But, in the SP programme of research, there has, so far, been no attempt to give the SPCM any kind of motivation (except the motivation `compress information'), or any kind of emotion. This is because of the belief that, in relation to the SP concepts and their development, it would be trying to run before we can walk. When the SP System is more mature, there will likely be a case for exploring how it may accommodate the complexities of motivations and emotions.

\section{Conclusion}\label{conclusion-section}

This paper describes several problems in AI research and how the {\em SP System} (Section \ref{intro-to-the-sp-system-section}) may solve them.

Most of the problems in AI research considered in this paper are described by leading researchers in AI, in interviews with science writer Martin Ford, and reported by him in his book {\em Architects of Intelligence} \cite{ford-2018}.

On the strength of evidence presented in this paper, there is clear potential for the SP System to help solve all the problems described in the paper. More generally, it appears that {\em the SP System provides a relatively firm foundation for the development of artificial general intelligence.} (Section \ref{introduction-section}).

The problems discussed in this paper, and how the SP System may solve them, are summarised as follows:

\paragraph{The need to bridge the divide between symbolic and non-symbolic kinds of knowledge representation and processing} (Section \ref{symbolic-non-symbolic-section}). The fact that people can deal effectively with symbolic systems such as natural languages and mathematics, and they can also do what appear to be non-symbolic things like recognising objects, playing tennis, and so on, points to the need for a system that can bridge those two styles of knowledge representation and processing. The SP System has clear potential in that area.

\paragraph{The tendency of deep neural networks to make large and unexpected errors in recognition} (Section \ref{dnns-easily-fooled-section}). The SP System appears to be entirely free from the tendency of DNNs to make large and unexpected errors in recognition.

\paragraph{The need to strengthen the representation and processing of natural languages} (Section \ref{rk-processing-of-nl-section}). Unlike DNNs, the SP System has strengths in the representation and processing of natural language with the kinds of syntactic structures that are recognised by linguists, with potential for the representation of semantic structures and their integration with syntax. It appears that such structures have psychological validity, and will probably be necessary for the achievement of human-like capabilities with natural languages.

\paragraph{The challenges of unsupervised learning} (Section \ref{challenge-of-unsupervised-learning-section}). Unlike most DNNs, learning in the SP System is entirely unsupervised. Although there are shortcomings as the system is now, it appears that the problems are soluble. This form of learning is prominent in the way people learn, and it may provide a foundation for other kinds of learning such as learning by being told, learning by imitation, learning via rewards and punishments (reinforcement learning), and so on.

\paragraph{The need for a coherent account of generalisation, under-generalisation, and over-generalisation} (Section \ref{generalisation-section}). The central role of IC in the SP system provides the basis for generalisation in accordance with coherent principles.

In brief, the theory of generalisation embodied in the SP System may be expressed quite simply: for a given body of information, {\bf I}, compress {\bf I} as much as possible, separate `encoding' from `grammar' and discard the encoding. Section \ref{generalisation-via-unsupervised-learning-section} describes how this works with the correction of over-generalisations, under-generalisations, and `dirty data', and Section \ref{generalisation-via-perception-section} describes how the principle would work with perception or related processes such as parsing.

Informal tests with the SPCM and earlier IC-based models of learning, suggest that these principles are sound.

\paragraph{How to learn usable knowledge from a single exposure or experience} (Section \ref{learning-from-a-single-experience-section}). Like people, and unlike DNNs, the SP System can learn useful knowledge from a single occurrence or experience. This is because: 1) the first stage in learning in the SP System is to take in information from the system's `environment' and to interpret it in terms of existing knowledge; and 2) without any further processing, that new knowledge can serve immediately in other AI functions such as reasoning, problem-solving, and so on. This contrasts with DNNs where much repetition is required before any new knowledge is usable.

\paragraph{How to achieve transfer learning (incorporating old knowledge in new)} (Section \ref{transfer-learning-section}). Transfer learning---meaning the use of old learning to facilitate later tasks---is prominent in human learning and is fundamental in the SP System. In this respect, the SP System contrasts sharply with DNNs. As noted in Section \ref{transfer-learning-section}, this capability is related to but distinct from `catastrophic forgetting' (Section \ref{catastrophic-forgetting-section}).

\paragraph{How to increase the speed of learning in AI systems, and how to reduce the demands of AI learning for large volumes of data, and for large computational resources} (Section \ref{comp-resources-speed-volumes-of-data-section}). Like people, and unlike DNNs, the SP System can demonstrate useful learning fast with relatively small demands for data and computational resources. This probably mainly because of the SP System's strengths in one-shot learning and transfer learning. With the learning of complex knowledge or skills such as a new natural language, the skills needed for gymnastics to a high level, and so on, the SP System, like people, is likely to need more time, because of the volume of knowledge and the complexity of the task of organising it efficiently.

\paragraph{The need for transparency in the representation and processing of knowledge} (Section \ref{transparency-in-rk-and-processing-section}). Unlike DNNs, the SP System is entirely transparent in how it represents knowledge, and it provides a full and comprehensible audit trail for all its processing.

\paragraph{How to achieve human-like capabilities in probabilistic reasoning} (Section \ref{probabilistic-reasoning-section}). As a by-product of its design, the SP System exhibits several forms of probabilistic reasoning, such as abductive reasoning, reasoning with discrimination networks and trees, nonmonotonic reasoning and more. It seems likely that capabilities like these will be needed in any system that aspires to AGI. As with the processing of natural language, there is clear potential for smooth integration of one aspect of AI---the SP System's strengths in probabilistic reasoning---with any or all of the SP System's other strengths in AI and the representation of knowledge.

\paragraph{The need to re-balance research towards top-down strategies} (Section \ref{top-down-strategy-section}). Unlike most research in AI today, the SP System has been developed via a top-down strategy, aiming for SIABC. This kind of top-down strategy seems to be essential if we are aiming for AGI. Owing to the overarching strategy adopted in the SP programme of research---seeking to simplify and integrate across a broad canvass---and owing to the SP System's favourable combination of conceptual Simplicity with descriptive and explanatory Power, it appears that {\em the SP System provides a relatively promising foundation for the development of artificial general intelligence.}.

\paragraph{How to minimise the risk of accidents with self-driving vehicles} (Section \ref{driverless-cars-section}). The complexities of driving a car seem to require human-like abilities in generalising its knowledge, with the correction of over- and under-generalisations. With the SP System, this may be done both via `generalisation via unsupervised learning' (Section \ref{generalisation-via-unsupervised-learning-section}) and `generalisation via perception' (Section \ref{generalisation-via-perception-section})

\paragraph{The need for strong compositionality in the structure of knowledge} (Section \ref{strong-compositionality-section}). Although DNNs appear to provide some approximations of the idea of one concept being composed of smaller elements, this idea is much more strongly developed in the SP System where it is entirely feasible to create class-inclusion hierarchies and part-whole hierarchies of any depth, and without the vagueness and ambiguity of DNN approximations of compositionality.

\paragraph{The challenges of commonsense reasoning and commonsense knowledge} (Section \ref{csrk-section}). Because commonsense reasoning and commonsense knowledge (CSRK) have developed as a distinct field within AI, it is discussed in a separate section, although the SP System's strengths in probabilistic reasoning Section \ref{probabilistic-reasoning-section} are part of its potential with CSRK. Other strengths of the SP System in that area are described.

\paragraph{Establishing the key importance of IC in AI research} (Section \ref{central-role-for-ic-section}). The SP System appears to be unique in employing IC as the basis for all aspects of intelligence and the representation of knowledge. There is much evidence for the importance of IC in HLPC \cite{sp-compression}.

\paragraph{Establishing the importance of a biological perspective in AI research} (Section \ref{biological-perspective-section}). Arguably, the SP System, with {\em SP-Neural}, has greater validity in terms of biology than DNNs, both in its organisation and in the central role of IC in how it works.

\paragraph{Establishing whether knowledge in the brain is represented in `distributed' or `localist' form} (Section \ref{distributed-localist-representations-section}). In contrast to DNNs, the SP System stores knowledge in an unambiguously `localist' form, meaning that each concept is represented by a small cluster of neurons in one location. In this respect it conforms to what appears to be the balance of evidence in favour of localist kinds of representation.

\paragraph{How to bypass the limited scope for adaptation in deep neural networks} (Section \ref{scope-for-adaptation-section}). The representation of knowledge with SP-patterns and with the SPMA construct, provides for much greater scope for adaptation than the layers of a DNN.

\paragraph{How to eliminate the problem of catastrophic forgetting} (Section \ref{catastrophic-forgetting-section}). The problem of catastrophic forgetting, where new knowledge wipes out old knowledge, is a significant problem with most DNNs. By contrast, the SP System is entirely free of the problem.

\paragraph{Overall conclusion}. Taken together, the evidence in these sections, and evidence of the versatility of the SP System (Appendices \ref{versatility-in-aspects of-intelligence-appendix}, \ref{versatility-in-reasoning-appendix}, \ref{versatility-in-representation-of-knowledge-appendix}, and \ref{versatility-in-representation-of-knowledge-appendix}), suggest that the SP System provides a firmer foundation for the development of human-like general AI (AGI, Section \ref{top-down-strategy-section}) than any alternative.

\section*{Acknowledgements}

I am grateful to anonymous reviewers for constructive comments on earlier drafts of this paper.

\begin{appendix}

\section{Mathematics incorporated in the SP Computer Model or contributing to its development}\label{mathematics-in-sp-system-appendix}

This appendix details mathematics incorporated in the SP Computer Model or contributing to its development. It is adapted with permission from \cite[Appendix A]{sp-compression}.

\subsection{Searching for repeating patterns}\label{searching_for_repeating_patterns_appendix}

At first sight, the process of searching for repeating patterns (Appendix \ref{ic-via-icmup-appendix}) is simply a matter of comparing one pattern with another to see whether they match each other or not. But there are, typically, many alternative ways in which patterns within a given body of information, {\bf I}, may be compared---and some are better than others.

We are interested in finding those matches between patterns that represent most redundancy and thus, via unification, yield most compression---and a little reflection shows that this is not a trivial problem \cite[Section 2.2.8.4]{wolff-2006}.

Maximising the amount of redundancy found means maximising $R$ where:

\begin{equation}
R = \sum_{i = 1}^{i = n} (f_i - 1) \cdot s_i,
\label{pattern_matching_equation}
\end{equation}

\noindent $f_i$ is the frequency of the $i$th member of a set of $n$ patterns and $s$ is its size in bits. Patterns that are both big and frequent are best. This equation applies irrespective of whether the patterns are coherent substrings or patterns that are discontinuous within {\bf I}.

Maximising $R$ means searching the space of possible unifications for the set of big, frequent patterns that gives the best value. For a sequence containing $N$ symbols, the number of possible subsequences (including single symbols and all composite patterns, both coherent and fragmented) is:

\begin{equation}
P = 2^{N} - 1.
\label{unifications_equation}
\end{equation}

The number of possible comparisons is the number of possible pairings of subsequences which is:

\begin{equation}
C = P(P - 1) / 2.
\label{comparisons_equation}
\end{equation}

For all except the very smallest values of $N$, the value of $P$ is very large and the corresponding value of $C$ is huge. In short, the abstract space of possible comparisons between patterns and thus the space of possible unifications is, in the great majority of cases, astronomically large.

Since the space is normally so large, it is not feasible to search it exhaustively. For that reason, it is normally necessary to use heuristic methods in searching---conducting the search in stages and discarding all but the best results at the end of each stage---and we must be content with answers that are ``reasonably good''.

Because it is not normally possible to use exhaustive search, we cannot normally guarantee to find the theoretically ideal answer. And, normally, we cannot know whether or not we have found that theoretically ideal answer.

\subsection{Information, compression of information, inductive inference and probabilities}\label{ic_inductive_inference_probabilies_appendix}

Solomonoff \cite{solomonoff-1964,solomonoff-1997} seems to have been one of the first people to recognise the close connection that exists between IC and {\em inductive inference}: predicting the future from the past, and calculating probabilities for such inferences. The connection between them---which may at first sight seem obscure---lies in the redundancy-as-repetition-of-patterns view of redundancy (Appendix \ref{ic-via-icmup-appendix}):

\begin{itemize}

\item Patterns that repeat within {\bf I} represent redundancy in {\bf I}, and IC can be achieved by reducing multiple instances of any pattern to one.

\item When we make inductive predictions about the future, we do so on the basis of repeating patterns. For example, the repeating pattern `Spring, Summer, Autumn, Winter' enables us to predict that, if it is Spring time now, Summer will follow.

\end{itemize}

Thus IC and inductive inference are closely related to concepts of frequency and probability. Here are some of the ways in which these concepts are related:

\begin{itemize}

\item Probability has a key role in Shannon's concept of information. In that perspective, the average quantity of information conveyed by one symbol in a sequence is:

    \begin{equation}
        H = - \sum_{i = 1}^{i = n} p_i \log p_i,
    \label{shannons_information_equation}
    \end{equation}

\noindent where $p_i$ is the probability of the $i$th type in the alphabet of $n$ available alphabetic symbol types. If the base for the logarithm is 2, then the information is measured in `bits'.

\item Measures of frequency or probability are central in techniques for economical coding such as the Huffman method \cite[Section 5.6]{cover-thomas-2006} or the Shannon-Fano-Elias method \cite[Section 5.9]{cover-thomas-2006}.

\item In the redundancy-as-repetition-of-patterns view of redundancy and IC, the frequencies of occurrence of patterns in {\bf I} is a main factor (with the sizes of patterns) that determines how much compression can be achieved.

\item Given a body of (binary) data that has been `fully' compressed (so that it may be regarded as random or nearly so), its absolute probability may be calculated as $p_{ABS} = 2^{-L}$, where $L$ is the length (in bits) of the compressed data.

\end{itemize}

Probability and IC may be regarded as two sides of the same coin. That said, they provide different perspectives on a range of problems. In this research, the IC perspective---with redundancy-as-repetition-of-patterns---seems to be more fruitful than viewing the same problems through the lens of probability. In the first case, one can see relatively clearly how compression may be achieved by the primitive operation of unifying patterns whereas these ideas are obscured when the focus is on probabilities.

\subsection{Random-dot stereograms}\label{random-dot_stereograms_appendix}

A particularly clear example of the kind of search described in Appendix \ref{ic-via-icmup-appendix} is what the brain has to do to enable one to see the figure in the kinds of random-dot stereogram described in \cite[Section 11]{sp-compression}.

In this case, assuming the left image has the same number of pixels as the right image, the size of the search space is:

\begin{equation}
S = P^2 / 2
\end{equation}

\noindent where $P$ is the number of possible patterns in each image, calculated in the same way as was described in Appendix \ref{searching_for_repeating_patterns_appendix}. The fact that the images are two dimensional needs no special provision because the original equations cover all combinations of atomic symbols.

For any stereogram with a realistic number of pixels, this space is very large indeed. Even with the very large processing power represented by the $10^{11}$ neurons in the brain, it is inconceivable that this space can be searched in a few seconds and to such good effect without the use of heuristic methods.

David Marr \cite[Chapter 3]{marr-1982} describes two algorithms that solve this problem. In line with what has just been said, both algorithms rely on constraints on the search space and both may be seen as incremental search guided by redundancy-related metrics.

\subsection{Coding and the evaluation of SP-multiple-alignments in terms of IC}\label{coding_sp-multiple-alignments_appendix}

Given an SP-multiple-alignment like one of the two shown in Figure \ref{parsing-kittens-figure} (Section \ref{demo-language-section}), one can derive a {\em code SP-pattern} from the SP-multiple-alignment in the following way:

\begin{enumerate}

\item Scan the SP-multiple-alignment from left to right looking for columns that contain an SP-symbol by itself, not aligned with any other symbol.

\item Copy these SP-symbols into a code pattern in the same order that they appear in the SP-multiple-alignment.

\end{enumerate}

\noindent The code SP-pattern derived in this way from the SP-multiple-alignment shown in Figure \ref{parsing-kittens-figure} is `\texttt{S PL 4 5 1 \#S}'. This is, in effect, a compressed representation of those symbols in the New pattern that form hits with Old symbols in the SP-multiple-alignment.

Given a code SP-pattern derived in this way, we may calculate a `compression difference' as:

\begin{equation}
CD = B_N - B_E
\label{CD_equation}
\end{equation}

\noindent or a `compression ratio' as:

\begin{equation}
CR = B_N / B_E,
\label{CR_equation}
\end{equation}

\noindent where $B_N$ is the total number of bits in those symbols in the New pattern that form hits with Old symbols in the SP-multiple-alignment, and $B_E$ is the total number of bits in the code SP-pattern (the `encoding') that has been derived from the SP-multiple-alignment as described above.

In each of these equations, $B_N$ is calculated as:

\begin{equation}
B_N = \sum_{i=1}^h C_i,
\label{BN_equation}
\end{equation}

\noindent where $C_i$ is the size of the code for $i$th symbol in a sequence, $H_1 ... H_h$, comprising those symbols within the New pattern that form hits with Old symbols within the SP-multiple-alignment (Appendix \ref{encoding_individual_symbols_appendix}).

$B_E$ is calculated as:

\begin{equation}
B_E = \sum_{i=1}^s C_i,
\label{BE_equation}
\end{equation}

\noindent where $C_i$ is the size of the code for $i$th symbol in the sequence of $s$ symbols in the code pattern derived from the SP-multiple-alignment (Appendix \ref{encoding_individual_symbols_appendix}).

\subsection{Encoding individual symbols}\label{encoding_individual_symbols_appendix}

The simplest way to encode individual symbols in the New pattern and the set of Old patterns in an SP-multiple-alignment is with a `block' code using a fixed number of bits for each symbol. But the SP Computer Model uses variable-length codes for symbols, assigned in accordance with the Shannon-Fano-Elias coding scheme \cite[Section 5.9]{cover-thomas-2006} so that the shortest codes represent the most frequent alphabetic symbol types and {\em vice versa}. Although this scheme is slightly less efficient than the well-known Huffman scheme, it has been adopted because it avoids some anomalous results that can arise with the Huffman scheme.

For the Shannon-Fano-Elias calculation, the frequency of each alphabetic symbol type ($f_{st}$) is calculated as:

\begin{equation}
f_{st} = \sum_{i=1}^P (f_i \times o_i)
\label{fst_equation}
\end{equation}

\noindent where $f_i$ is the (notional) frequency of the $i$th pattern in the collection of Old SP-patterns (the {\em grammar}) used in the creation of the given SP-multiple-alignment, $o_i$ is the number of occurrences of the given symbol in the $i$th SP-pattern in the grammar and $P$ is the number of SP-patterns in the grammar.

\subsection{Calculation of probabilities associated with any given SP-multiple-alignment}\label{probabilities_calculation_appendix}

As may be seen in \cite[Chapter 7]{wolff-2006}, the formation of SP-multiple-alignments in the SP framework supports a variety of kinds of probabilistic reasoning. The core idea is that any Old symbol in a SP-multiple-alignment that is {\em not} aligned with a New symbol represents an inference that may be drawn from the SP-multiple-alignment. This section describes how absolute and relative probabilities for such inferences may be calculated.

\subsubsection{Absolute probabilities}

Any sequence of $L$ symbols, drawn from an alphabet of $|A|$ alphabetic types, represents one point in a set of $N$ points where $N$ is calculated as:

\begin{equation}
N = |A|^L.
\label{N_equation}
\end{equation}

\noindent {\em If we assume that the sequence is random or nearly so}, which means that the $N$ points are equi-probable or nearly so, the probability of any one point (which represents a sequence of length $L$) is close to:

\begin{equation}
p_{ABS} = |A|^{-L}.
\label{pABS_equation}
\end{equation}

\noindent In the SP Computer Model, the value of $|A|$ is $2$.

This equation may be used to calculate the absolute probability of the code, $C$, derived from the SP-multiple-alignment as described in Appendix \ref{coding_sp-multiple-alignments_appendix}. $p_{ABS}$ may also be regarded as the absolute probability of any inferences that may be drawn from the SP-multiple-alignment as described in \cite[Section 7.2.2]{wolff-2006}.

\subsubsection{Relative probabilities}

The absolute probabilities of SP-multiple-alignments, calculated as described in the last subsection, are normally very small and not very interesting in themselves. From the standpoint of practical applications, we are normally interested in the {\em relative} values of probabilities, not their {\em absolute} values.

The procedure for calculating relative values for probabilities ($p_{REL}$) is as follows:

\begin{enumerate}

\item For the SP-multiple-alignment which has the highest $CD$ (which we shall call the {\em reference SP-multiple-alignment}), identify the symbols from New which are encoded by the SP-multiple-alignment. We will call these symbols the {\em reference set of symbols in New}.

\item Compile a {\em reference set of SP-multiple-alignments} which includes {\em the SP-multiple-alignment with the highest $CD$ and all other SP-multiple-alignments (if any) which encode exactly the reference set of symbols from New, neither more nor less}.

\item The SP-multiple-alignments in the reference set are examined to find and remove any rows which are redundant in the sense that all the symbols appearing in a given row also appear in another row in the same order.\footnote{If Old is well compressed, this kind of redundancy amongst the rows of a SP-multiple-alignment should not appear very often.} Any SP-multiple-alignment which, after editing, matches another SP-multiple-alignment in the set is removed from the set.

\item Calculate the sum of the values for $p_{ABS}$ in the reference set of SP-multiple-alignments:

\begin{equation}
p_{A\_SUM} = \sum_{i = 1}^{i = R} p_{ABS_i}
\label{pA_SUM_equation}
\end{equation}

\noindent where $R$ is the size of the reference set of SP-multiple-alignments and $p_{ABS_i}$ is the value of $p_{ABS}$ for the $i$th SP-multiple-alignment in the reference set.

\item For each SP-multiple-alignment in the reference set, calculate its relative probability as:

\begin{equation}
p_{REL_i} = p_{ABS_i} / p_{A\_SUM}.
\label{pREL_equation}
\end{equation}

\end{enumerate}

The values of $p_{REL}$ calculated as just described seem to provide an effective means of comparing the SP-multiple-alignments in the reference set. Normally, this will be those SP-multiple-alignments which encode the same set of symbols from New as the SP-multiple-alignment which has the best overall $CD$.

\subsection{Sifting and sorting of SP-patterns in unsupervised learning in the SP System}\label{sifting_and_sorting_appendix}

In the process of unsupervised learning in the SP System (Section \ref{challenge-of-unsupervised-learning-section}, Appendix \ref{sp-unsupervised-learning-appendix}, and \cite[Chapter 9]{wolff-2006}), which starts with a set of New SP-patterns, there is a process of sifting and sorting Old SP-patterns that are created by the SP System to develop one or more alternative collections of Old SP-patterns ({\em grammars}), each one of which scores well in terms of its capacity for the economical encoding of the given set of New SP-patterns.

When all the New SP-patterns have been processed in this way, there is a set $A$ of full SP-multiple-alignments, divided into $b_1 ... b_m$ disjoint subsets, one for each SP-pattern from the given set of New SP-patterns. From these SP-multiple-alignments, the program computes the frequency of occurrence of each of the $p_1 ... p_n$ Old SP-patterns as:

\begin{equation}
f_i = \sum_{j = 1}^{j = m} max(p_i, b_j)
\label{frequency_of_patterns_equation}
\end{equation}

\noindent where $max(p_i, b_j)$ is the maximum number of times that $p_i$ appears in any {\em one} of the SP-multiple-alignment in the subset $b_j$.

The program also compiles an alphabet of the alphabetic symbol types, $s_1 ... s_r$, in the Old SP-patterns and, following the principles just described, computes the frequency of occurrence of each alphabetic symbol type as:

\begin{equation}
F_i = \sum_{j = 1}^{j = m} max(s_i, b_j)
\label{frequency_of_symbol_types_equation}
\end{equation}

\noindent where $max(s_i, b_j)$ is the maximum number of times that $s_i$ appears in any {\em one} SP-multiple-alignment in subset $b_j$. From these values, the encoding cost of each alphabetic symbol type is computed using the Shannon-Fano-Elias method as before \cite[Section 5.9]{cover-thomas-2006}.

In the process of building alternative grammars, the tree of such alternatives is pruned periodically to keep it within reasonable bounds. Values for $G$, $E$ and $(G + E)$ (which we will refer to as $T$) are calculated for each grammar and, at each stage, grammars with high values for $T$ are eliminated.

For a given grammar comprising SP-patterns $p_1 ... p_g$, the value of $G$ is calculated as:

\begin{equation}
G = \sum_{i=1}^{i=g}(\sum_{j=1}^{j=L_i}s_j)
\label{size_of_grammar_equation}
\end{equation}

\noindent where $L_i$ is the number of symbols in the $i$th SP-pattern and $s_j$ is the encoding cost of the $j$th SP-symbol in that SP-pattern.

Given that each grammar is derived from a set $a_1 ... a_n$ of SP-multiple-alignments (one SP-multiple-alignment for each pattern from New), the value of $E$ for the grammar is calculated as:

\begin{equation}
E = \sum_{i=1}^{i=n}e_i
\label{size_of_encoded_data_equation}
\end{equation}

\noindent where $e_i$ is the size, in bits, of the code SP-pattern derived from the $i$th SP-multiple-alignment.

\subsection{Finding good matches between two sequences of symbols}\label{finding_good_matches_appendix}

At the heart of the SP Computer Model is a process for finding good matches between two sequences of symbols, outlined in Appendix \ref{spma-creation-appendix} and described quite fully in \cite[Appendix A]{wolff-2006}. What has been developed is a version of dynamic programming with the advantage that it can find two or more good matches between sequences, not just one good match.

The search process uses a measure of probability, $p_n$, as its metric. This metric provides a means of guiding the search which is effective in practice and appears to have a sound theoretical basis. To define $p_n$ and to justify it theoretically, it is necessary first to define the terms and variables on which it is based:

\begin{itemize}

\item A sequence of matches between two sequences, sequence1 and sequence2, is called a `hit sequence'.

\item For each hit sequence $h_1 ... h_n$, there is a corresponding series of {\em gaps}, $g_1 ... g_n$. For any one hit, the corresponding gap is $g = g_q + g_d$, where $g_q$ is the number of unmatched characters in the query between the query character for the given hit in the series and the query character for the immediately preceding hit; and $g_d$ is the equivalent gap in the database, $g_1$ is taken to be 0.

\item $A$ is the size of the {\em alphabet} of symbol types used in sequence1 and sequence2.

\item $p_1$ is the probability of a match between any one symbol in sequence1 and any one symbol in sequence2 on the null hypothesis that all hits are equally probable at all locations. Its value is calculated as: $p_1 = 1 / A$.

\end{itemize}

Using these definitions, the probability of any hit sequence of length $n$ is calculated as:

\[p_n = \prod_{i=1}^{i=n}(1 - (1 - p_1)^{g_i + 1}), \hspace{3mm} g_1 = 0\].

With this equation, is relatively easy to calculate the probability of the hit sequence up to and including any hit by using the stored value of the hit sequence up to and including the immediately preceding hit.

\section{Outline of the SP Theory of Intelligence and its realisation in the SPCM}\label{outline-of-sp-system-appendix}

\sloppy The {\em SP System}---meaning the {\em SP Theory of Intelligence} and its realisation in the {\em SPCM} is the product of a lengthy programme of research, from about 1987 to now with a break between early 2006 and late 2012. This programme of research has included the creation and testing of many versions of the SPCM. A major discovery has been the concept of {\em SP-multiple-alignment} and its versatility in many aspects of intelligence (Appendix \ref{sp-multiple-alignment-appendix}).

\subsection{Aiming for a favourable combination of conceptual {\em Simplicity} with descriptive or explanatory {\em Power}}\label{simplicity-power-appendix}

{\em The overarching goal of the SP programme of research is to simplify and integrate observations and concepts across AI, mainstream computing, mathematics, and human learning, perception, and cognition} (SIABC). In effect, this means developing concepts that combine conceptual {\em Simplicity} with high levels of descriptive or explanatory {\em Power}. This in turn means the same as IC by increasing the {\em simplicity} of a body of information {\bf I}, by the removal of {\em redundancy} from {\bf I}, whilst retaining as much as possible of its non-redundant, expressive {\em power}.

As readers may guess, the Simplicity and Power concepts, which apply to both the aims of the research and the workings of the SPCM, are the reasons for the name `SP'.

Despite its ambition, the simplicity-with-power objective has been largely met. This is because the SP System, which is largely the simple but powerful concept of SPMA (Appendix \ref{sp-multiple-alignment-appendix}), with relatively simple processes for unsupervised learning (Appendix \ref{sp-unsupervised-learning-appendix}), has strengths and potential across diverse aspects of intelligence and the representation of diverse kinds of knowledge (Appendix \ref{sp-str-pot-appendix}).

\subsection{High level view of the SP System}\label{high-level-view-of-sp-appendix}

\begin{figure}[!htbp]
\centering
\includegraphics[width=0.7\textwidth]{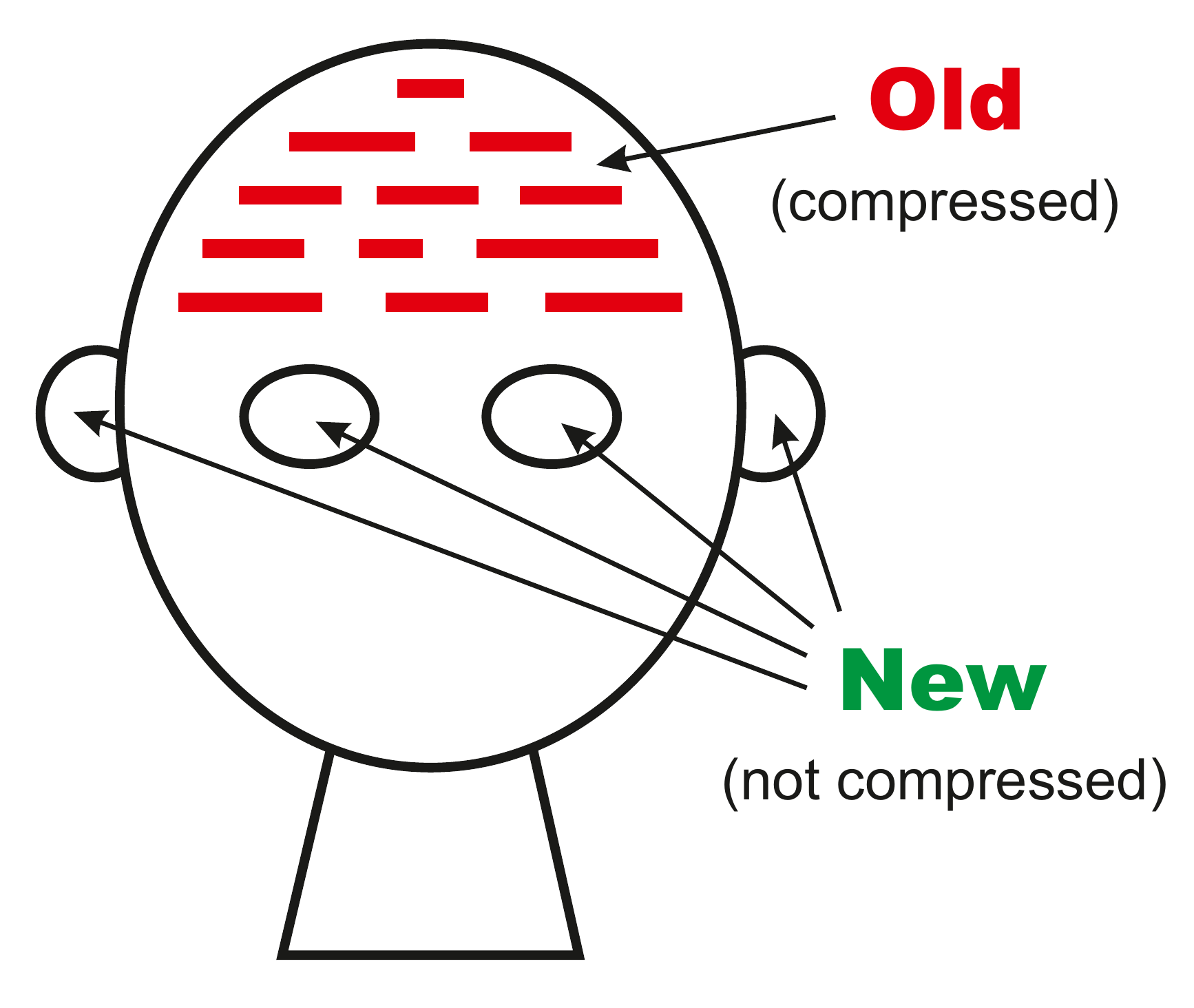}
\caption{Schematic representation of the SP System from an `input' perspective. Reproduced, with permission, from Figure 1 in \protect\cite{sp-extended-overview}.}
\label{sp-input-perspective-figure}
\end{figure}

In broad terms, the SP System is a brain-like system that takes in {\em New} information through its senses and stores some or all of it as {\em Old} information that is compressed, as shown schematically in Figure \ref{sp-input-perspective-figure}.

In the SP System, all kinds of knowledge are represented with {\em SP-patterns}, where each such SP-pattern is an array of atomic {\em SP-symbols} in one or two dimensions. An SP-symbol is simply a `mark' that can be matched with any other SP-symbol to determine whether it is the same or different.

At present, the SPCM works only with one-dimensional SP-patterns but it is envisaged that it will be generalised to work with two-dimensional SP-patterns as well.

An important point here is that, {\em because the SP System cannot yet process two-dimensional SP-patterns, it cannot demonstrate directly the way in which images would be processed in the system when it is more mature. However, there are many aspects of image processing that can be demonstrated with 2D SP-patterns.}

\subsection{Information compression in the SP System via the matching and unification of patterns}\label{ic-via-icmup-appendix}

A central idea in the SP System is that all kinds of processing would be achieved via IC. Evidence for the importance of IC in HLPC is described in papers by pioneers in this area including \cite{attneave-1954,barlow-1959,barlow-1969}, and in a more recent review \cite{sp-compression}.

In the development of the SP System, it has proved useful to understand IC as a process of searching for patterns that match each other and the merging or `unifying' patterns that are the same. The expression `information compression via the matching and unification of patterns' may be abbreviated as `ICMUP'.

More specifically IC in the SP System is achieved largely via the creation of SPMAs (Appendix \ref{sp-multiple-alignment-appendix}) and, via unsupervised learning (with SPMAs playing an important role), the creation of SP-grammars (Appendix \ref{sp-unsupervised-learning-appendix}).

In terms of SP Theory, the emphasis on IC in the SP System accords with research in the tradition of Minimum Length Encoding (see, for example, \cite{li-vitanyi-2019}), with the qualification that most research relating to MLE assumes that the concept of a universal Turing machine provides the foundation for theorising, whereas the SP System is itself a theory of computing \cite[Section II-C]{sp-alternatives} founded on concepts of ICMUP and SPMA.

For the time being, the SPCM has been and now is being developed for lossless IC, but there may be a case in the future for exploring potential benefits if the SPCM provides an option for lossy IC.

\subsection{SP-multiple-alignment}\label{sp-multiple-alignment-appendix}

A central idea in the SP System, is the simple but powerful concept of SPMA, borrowed and adapted from the concept of `multiple sequence alignment' in bioinformatics.

SPMA is the last of seven variants of ICMUP described in \cite[Section 5]{sp-micmup}. It may be seen to be a generalisation of the other six variants \cite[Section 5.7]{sp-micmup}.

Within the SP System, the SPMA concept is largely responsible for the AI-related strengths and potential of the SP System as outlined in Appendix \ref{sp-str-pot-appendix}.

The versatility of the SP System may also be seen in its several potential areas of application summarised in Appendix \ref{benefits-and-applications-appendix}.

Bearing in mind that it is just as bad to underplay the strengths and potential of a system as it is to oversell its strengths and potential, it seems fair to say that {\em the concept of SP-multiple-alignment may prove to be as significant for an understanding of `intelligence' as is DNA for biological sciences. It may prove to be the `double helix' of intelligence}.

\subsubsection{Multiple sequence alignments}

As mentioned above, the SPMA concept has been borrowed and adapted from the concept of `multiple sequence alignment' in bioinformatics. Figure \ref{ma-dna-figure} shows an example. Here, there are five DNA sequences which have been arranged alongside each other, and then, by judicious `stretching' of one or more of the sequences in a computer, symbols that match each other across two or more sequences have been brought into line.

\begin{figure}[!htbp]
\fontsize{10.00pt}{12.00pt}
\centering
{\bf
\begin{BVerbatim}
  G G A     G     C A G G G A G G A     T G     G   G G A
  | | |     |     | | | | | | | | |     | |     |   | | |
  G G | G   G C C C A G G G A G G A     | G G C G   G G A
  | | |     | | | | | | | | | | | |     | |     |   | | |
A | G A C T G C C C A G G G | G G | G C T G     G A | G A
  | | |           | | | | | | | | |   |   |     |   | | |
  G G A A         | A G G G A G G A   | A G     G   G G A
  | |   |         | | | | | | | |     |   |     |   | | |
  G G C A         C A G G G A G G     C   G     G   G G A
\end{BVerbatim}
}
\caption{A `good' multiple sequence alignment amongst five DNA sequences.}
\label{ma-dna-figure}
\end{figure}

A `good' multiple sequence alignment, like the one shown, is one with a relatively large number of matching symbols from row to row. The process of discovering a good multiple sequence alignment is normally too complex to be done by exhaustive search, so heuristic methods are needed, building multiple sequence alignments in stages and, at each stage, selecting the best partial structures for further processing.

Some people may argue that the combinational explosion with this kind of problem, and the corresponding computational complexity, is so large that there is no practical way of dealing with it. In answer to that objection, there are several multiple sequence alignment programs used in bioinformatics---such as `Clustal Omega', `Kalign', and `MAFFT'\footnote{Provided as online services by the European Bioinformatics Institute (see \href{www.ebi.ac.uk/Tools/msa/}{https://www.ebi.ac.uk/Tools/msa/}).}---which produce results that are good enough for practical purposes.

This relative success is achieved via the use of heuristic methods that conduct the search for good structures in stages, discarding all but the best alignments at the end of each stage. With these kinds of methods, reasonably good results may be achieved but normally they cannot guarantee that the best possible result has been found.

\subsubsection{How SP-multiple-alignments are created}\label{spma-creation-appendix}

Figure \ref{parsing-kittens-figure} (in Section \ref{rk-processing-of-nl-section}) shows an example of an SPMA, superficially similar to the one in Figure \ref{ma-dna-figure}, except that the sequences are called {\em SP-patterns}, the SP-pattern in row 0 is New information and each of the remaining SP-patterns, one per row, is an Old SP-pattern, selected from a relatively large pool of such SP-patterns. A `good' SPMA is one which allows the New SP-pattern to be encoded economically in terms of the Old SP-patterns.

In this example, the New SP-pattern (in row 0) is a sentence and each of the remaining SP-patterns represents a grammatical category, where `grammatical categories' include words. The overall effect of the SPMA in this example is the parsing of a sentence (`\texttt{f o r t u n e f a v o u r s t h e b r a v e}') into its grammatical parts and sub-parts.

As with multiple sequence alignments, it is almost always necessary to use heuristic methods to achieve useful results without undue computational demands. The use of heuristic methods helps to ensure that computational complexities in the SP System are within reasonable bounds \cite[Sections A.4, 3.10.6 and 9.3.1]{wolff-2006}. Each SP-multiple-alignment is built up progressively, starting with a process of finding good alignments between pairs of SP-patterns. At the end of each stage, SP-multiple-alignments that score well in terms of IC are retained and the rest are discarded. There is more detail in \cite[Section 4]{sp-extended-overview} and \cite[Sections 3.4 and 3.5]{wolff-2006}.

In the SPCM, the size of the memory available for searching may be varied, which means in effect that the scope for backtracking can be varied. When the scope for backtracking is increased, the chance of the program getting stuck on a `local peak' (or `local minimum') in the search space is reduced.

Contrary to the impression that may be given by Figure \ref{parsing-kittens-figure}, the SPMA concept is very versatile and is largely responsible for the strengths and potential of the SP System, as described in Appendix \ref{sp-str-pot-appendix}.

\subsubsection{The SP System is {\em quite different} from a deep neural network}\label{sp-is-not-dnn-appendix}

The several levels in an SPMA may give the impression that the SP System in its structure and workings is simply a variant of the structure and workings of a DNN. This is entirely false.

In DNNs, the layers are provided at the beginning of processing and do not change except in the strengthening of links between neurons. By contrast, the SP System stores its knowledge in the form of SP-patterns, and those SP-patterns become the rows in each of a multitude of different SPMAs, each of which contains its own distinctive array of SP-patterns, normally a unique set of SP-patterns but sometimes two or more sets are the same but with different alignments.
Also, IC is of central importance in the SP System by contrast with most DNNs in which IC has little or no role \cite{schmidhuber-2015}.

Distinctive features and advantages of the SP System are described more fully in \cite{sp-alternatives}.

\subsection{Unsupervised learning in the SP System}\label{sp-unsupervised-learning-appendix}

In the SP System, learning is `unsupervised', deriving structures from incoming sensory information without the need for any kind of `teacher', or anything equivalent ({\em cf.} \cite{gold-1967}).

Unsupervised learning in the SP System is quite different from `Hebbian' learning via the gradual strengthening or weakening of neural connections (Section {challenge-of-unsupervised-learning-section}), variants of which are the mainstay of learning in DNNs. In the SP System, unsupervised learning incorporates the building of SPMAs but there are other processes as well.

\subsubsection{The creation of Old SP-patterns}\label{creation-of-sp-patterns-appendix}

In brief, the system creates Old SP-patterns from complete New SP-patterns and also from partial matches between New and Old SP-patterns. All learning in the SP System starts with the taking in of information from the environment:

\begin{itemize}

    \item If that information is the same as one or more Old SP-patterns, then the frequency of the one SP-pattern is increased, or frequencies of the two or more SP-patterns are increased.

    \item If that information is entirely new, `ID' SP-symbols\footnote{Appendix \ref{high-level-view-of-sp-appendix}} are added at the beginning and end of the pattern so that it becomes an SP-pattern. Then it is added directly to the store of Old SP-patterns.

    \item If partial matches can be made between the newly-received information and one or more of the stored Old SP-patterns, then each of the parts that match, and each of the parts that don't match, are made into SP-patterns by the addition of ID SP-symbols at the beginning and end, and the newly-created SP-patterns are added to the store of Old SP-patterns.

\end{itemize}

\subsubsection{The creation of SP-grammars}\label{creation-of-sp-grammars-appendix}

With a given body of New SP-patterns, the system processes them as just sketched, and then searches for one or two `good' {\em SP-grammars}, where an SP-grammar is a collection of Old SP-patterns, and it is `good' if it is effective in the economical encoding of the original set of New SP-patterns, where that economical encoding is achieved via SPMA.

As with the building of SPMAs, the process of creating good grammars is normally too complex to be done by exhaustive search so heuristic methods are needed. This means that the system builds SP-grammars incrementally and, at each stage, it discards all but the best SP-grammars.

As with the building of SPMAs, the use of heuristic methods helps to ensure that computational complexities in the SP System are within reasonable bounds \cite[Sections A.4, 3.10.6 and 9.3.1]{wolff-2006}.

The SPCM has already demonstrated an ability to learn generative grammars from unsegmented samples of English-like artificial languages, including segmental structures, classes of structure, and abstract patterns, and to do this in an `unsupervised' manner (\cite[Section 5]{sp-extended-overview}, \cite[Chapter 9]{wolff-2006}).

\subsubsection{Shortcomings in unsupervised learning in the SP System}\label{shortcomings-sp-learning-appendix}

But there are (at least) two shortcomings in the system \cite[Section 3.3]{sp-extended-overview}: it cannot learn intermediate levels of structure or discontinuous dependencies in grammar, although the SPMA framework can accommodate structures of those kinds. It appears that those two problems may be overcome and that their solution would greatly enhance the capabilities of the SPCM in unsupervised learning.

\subsection{The probabilistic nature of the SP System}\label{probabilistic-nature-of-sp-appendix}

Owing to the intimate relation that is known to exist between IC and concepts of probability \cite{solomonoff-1964,solomonoff-1997}, and owing to the fundamental role of IC in the workings of the SP System, the system is inherently probabilistic (\cite[Section 4.4]{sp-extended-overview}, \cite[Section 3.7]{wolff-2006}).

That said, it appears to be possible to imitate the all-nothing-nature of conventional computing systems via the use of data where all the probabilities yielded by the system are at or close to 0 or 1.

Because of the probabilistic nature of the SP System, it lends itself to the modelling of HLPC because of the prevalence of uncertainties in that domain. Also, the SP System sits comfortably within AI because of the probabilistic nature of most systems in AI, at least in more recent work in that area.

An advantage of the SP System in those areas is that it is relatively straightforward to calculate absolute or conditional probabilities for results obtained in, for example, different kinds of reasoning (Appendix \ref{versatility-in-reasoning-appendix}).

The very close connection that exists between IC and concepts of probability may suggest that there is nothing to choose between them. But \cite[Section 8.2]{sp-micmup} argues that, in research on aspects of AI and HLPC, there are reasons to regard IC as more fundamental than probability and a better starting point for theorising.

\subsection{Two high-level mechanisms for IC in the SP System, and their functions}\label{two-mechanisms-for-ic-appendix}

The two main mechanisms for IC in the SP System, both of which incorporate ICMUP (Appendix \ref{ic-via-icmup-appendix}), are as follows, each one with details of its function or functions:

\begin{enumerate}

    \item {\em The building of SP-multiple-alignments}. The process of building SPMAs achieves compression of New information. At the same time it may achieve any or all of the following functions described in \cite[Chapters 5 to 8]{wolff-2006} and \cite[Sections 7 to 12]{sp-extended-overview}, with potential for more:

        \begin{enumerate}

            \item The parsing of natural language (which is quite well developed); and understanding of natural language (which is only at a preliminary stage of development).

            \item Pattern recognition which is robust in the face of errors of omission, commission, or substitution; and pattern recognition at multiple levels of abstraction.

            \item Information retrieval which is robust in the face of errors of omission, commission, or substitution.

            \item Several kinds of probabilistic reasoning, as summarised in Appendix \ref{versatility-in-reasoning-appendix}.

            \item Planning such as, for example, finding a flying route between London and Beijing.

            \item Problem solving such as solving the kinds of puzzle that are popular in IQ tests.

        \end{enumerate}

        The building of SPMAs is also part of the process of unsupervised learning, next.

    \item {\em Unsupervised learning}. Unsupervised learning, outlined in Appendix \ref{sp-unsupervised-learning-appendix}, means the creation of one or two {\em SP-grammars} which are collections of SP-patterns which are effective in the economical encoding of a given set of New SP-patterns.

\end{enumerate}

\subsection{SP-Neural}\label{sp-neural-appendix}

A potentially useful feature of the SP System is that it is possible to see how abstract constructs and processes in the system may be realised in terms of neurons and their interconnections. This is the basis for {\em SP-Neural}, a `neural' version of the SP System, described in \cite{spneural-2016}.

\begin{figure}[!htbp]
\centering
\includegraphics[width=0.9\textwidth]{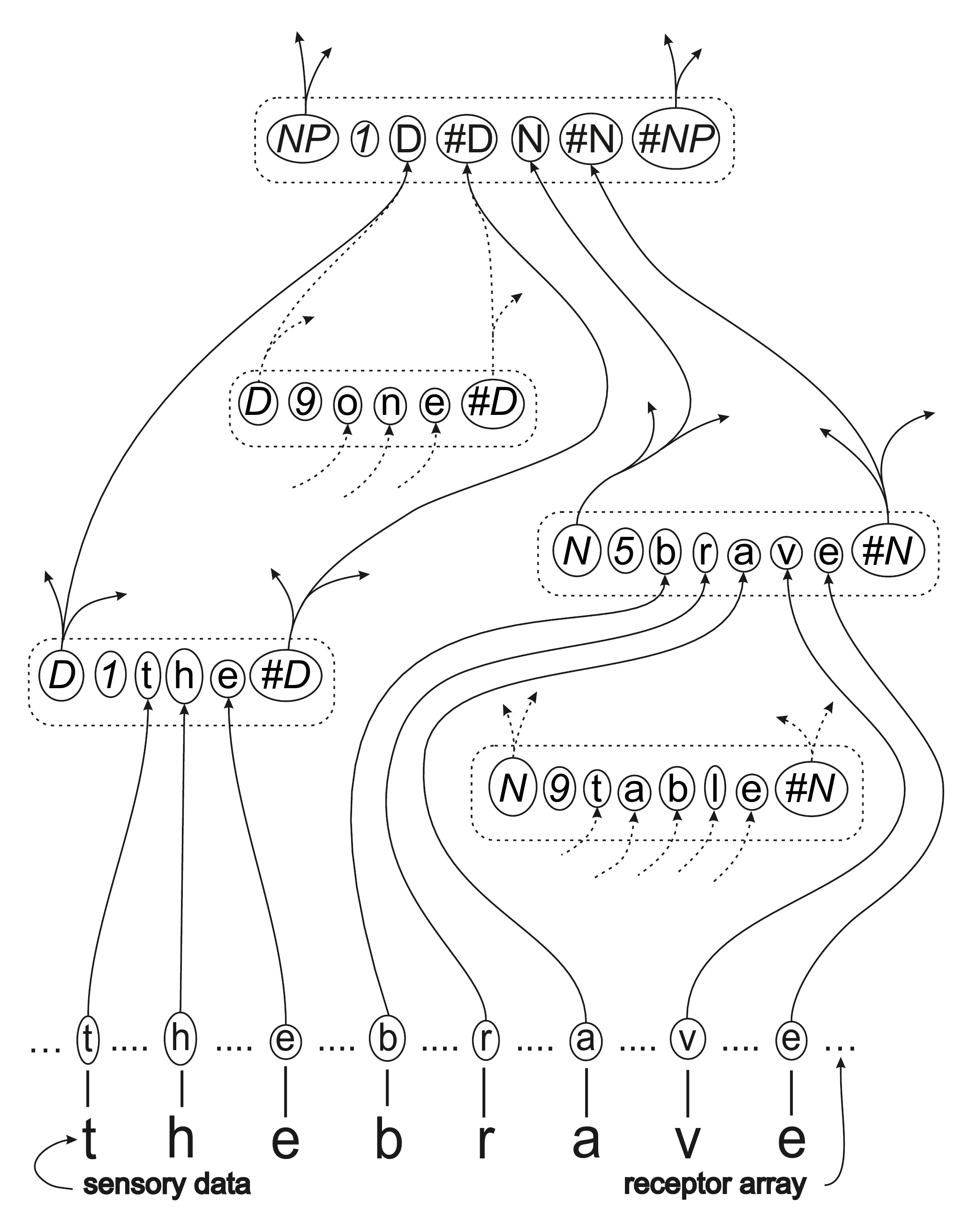}
\caption{A schematic representation of a partial SPMA in SP-Neural, as discussed in \cite[Section 4]{spneural-2016}. Each broken-line rectangle with rounded corners represents a {\em pattern assembly}---corresponding to an SP-pattern in the main SP Theory of Intelligence; each character or group of characters enclosed in a solid-line ellipse represents a {\em neural symbol} corresponding to an SP-symbol in the main SP Theory of Intelligence; the lines between pattern assemblies represent nerve fibres with arrows showing the direction in which impulses travel; neural symbols are mainly symbols from linguistics such as `\texttt{NP}' meaning `noun phrase, `\texttt{D}' meaning a `determiner', `\texttt{\#D}' meaning the end of a determiner, `\texttt{\#NP}' meaning the end of a noun phrase, and so on.}
\label{the-brave-neural-figure}
\end{figure}

The concept of an SP-symbol may realised as a {\em neural symbol} comprising a single neuron or, more likely, a small cluster of neurons. An SP-pattern maps quite well on to the concept of a {\em pattern assembly} comprising a group of inter-connected SP-symbols. And an SPMA may be realised in terms of pattern assemblies and their interconnections, as illustrated in Figure \ref{the-brave-neural-figure}.

In this connection, it is relevant to mention that the SP System, in both its abstract and neural forms, is quite different from DNNs \cite{schmidhuber-2015} and has substantial advantages compared with such systems, as described in several sections in this paper and in \cite[Section V]{sp-alternatives}.

\subsection{Generalising the SP System for two-dimensional SP-patterns, both static and moving}

This brief description of the SP System and how it works may have given the impression that it is intended to work entirely with sequences of SP-symbols, like multiple sequence alignments in bioinformatics. But it is envisaged that, in future development of the system, two-dimensional SP-patterns will be introduced, with potential to represent and process such things as photographs and diagrams, and structures in three dimensions as described in \cite[Section 6.1 and 6,2]{sp-vision}, and procedures that work in parallel as described in \cite[Sections V-G, V-H, and V-I, and C]{sp-autonomous-robots}.

{\em It is envisaged that, at some stage, the SP System will be generalised to work with two-dimensional `frames' from films or videos, and the sequencing needed to represent motion, and eventually the information needed to represent 3D bodies in motion, as in a 3D film.} These developments are needed for the processing of static and moving images. Nevertheless, quite a lot can be learned with the current SP Computer Model about the the processing of images.

\subsection{Strengths and potential of the SP System in intelligence-related functions and applications}\label{sp-str-pot-appendix}

The strengths and potential of the SP System are summarised in the subsections that follow. Further information may be found in \cite[Sections 5 to 12]{sp-extended-overview}, \cite[Chapters 5 to 9]{wolff-2006}, \cite{sp-alternatives}, and in other sources referenced in the subsections that follow.

In view of the relative Simplicity of the SP System, the strengths and potential of the system summarised here mean that the system combines relative Simplicity with relatively high levels of descriptive and explanatory Power (Appendix \ref{simplicity-power-appendix}).

\subsubsection{Versatility in aspects of intelligence}\label{versatility-in-aspects of-intelligence-appendix}

The SP System has strengths and potential in the `unsupervised' learning of new knowledge. As noted in Appendix \ref{sp-unsupervised-learning-appendix}, this is an aspect of intelligence in the SP System that is different from others because it is not a by-product of the building of multiple alignments but is, instead, achieved via the creation of {\em grammars}, drawing on information within SPMAs.

Other aspects of intelligence where the SP System has strengths or potential are modelled via the building of SPMAs. These other aspects of intelligence include: the analysis and production of natural language; pattern recognition that is robust in the face of errors in data; pattern recognition at multiple levels of abstraction; computer vision \cite{sp-vision}; best-match and semantic kinds of information retrieval; several kinds of reasoning (next subsection); planning; and problem solving.

\subsubsection{Versatility in reasoning}\label{versatility-in-reasoning-appendix}

Kinds of reasoning exhibited by the SP System include: one-step `deductive' reasoning; chains of reasoning; abductive reasoning; reasoning with probabilistic networks and trees; reasoning with `rules'; nonmonotonic reasoning and reasoning with default values; Bayesian reasoning with `explaining away'; causal reasoning; reasoning that is not supported by evidence; the inheritance of attributes in class hierarchies; and inheritance of contexts in part-whole hierarchies. Where it is appropriate, probabilities for inferences may be calculated in a straightforward manner (\cite[Section 3.7]{wolff-2006}, \cite[Section 4.4]{sp-extended-overview}).

There is also potential in the system for spatial reasoning \cite[Section IV-F.1]{sp-autonomous-robots}, and for what-if reasoning \cite[Section IV-F.2]{sp-autonomous-robots}.

It seems unlikely that the features of intelligence mentioned above are the full extent of the SP System's potential to imitate what people can do. The close connection that is known to exist between IC and concepts of probability (Appendix \ref{probabilistic-nature-of-sp-appendix}), the central role of IC in the SPMA framework, and the versatility of the SPMA framework in aspects of intelligence suggest that there are more insights to come.

As noted in Appendix \ref{probabilistic-nature-of-sp-appendix}, the probabilistic nature of the SP System makes it relatively straightforward to calculate absolute or conditional probabilities for results from the system, as for example in its several kinds of reasoning, most of which would naturally be classed as probabilistic.

\subsubsection{Versatility in the representation of knowledge}\label{versatility-in-representation-of-knowledge-appendix}

Although SP-patterns are not very expressive in themselves, they come to life in the SPMA framework. Within that framework, they may serve in the representation of several different kinds of knowledge, including: the syntax of natural languages; class-inclusion hierarchies (with or without cross classification); part-whole hierarchies; discrimination networks and trees; if-then rules; entity-relationship structures \cite[Sections 3 and 4]{sp-intelligent-database}; relational tuples ({\em ibid}., Section 3), and concepts in mathematics, logic, and computing, such as `function', `variable', `value', `set', and `type definition' (\cite[Chapter 10]{wolff-2006}, \cite[Section 6.6.1]{sp-benefits-apps}, \cite[Section 2]{sp-software-engineering}).

As previously noted, the addition of two-dimensional SP patterns to the SPCM is likely to expand the representational repertoire of the SP System to structures in two-dimensions and three-dimensions, and the representation of procedural knowledge with parallel processing.

As with the SP System's generality in aspects of intelligence, it seems likely that the SP System is not constrained to represent only the forms of knowledge that have been mentioned. The generality of IC as a means of representing knowledge in a succinct manner, the central role of IC in the SPMA framework, and the versatility of that framework in the representation of knowledge, suggest that the SP System may prove to be a means of representing {\em all} the kinds of knowledge that people may work with.

\subsubsection{The seamless integration of diverse aspects of intelligence, and diverse kinds of knowledge, in any combination}\label{seamless-integration-appendix}

An important third feature of the SP System, alongside its versatility in aspects of intelligence and its versatility in the representation of diverse kinds of knowledge, is that {\em there is clear potential for the SP System to provide SI.} This is because diverse aspects of intelligence and diverse kinds of knowledge all flow from a single coherent and relatively simple source: the SPMA framework.

It appears that SI is {\em essential} in any artificial system that aspires to the fluidity, versatility and adaptability of the human mind.

\begin{figure*}[!hbt]
\centering
\includegraphics[width=0.9\textwidth]{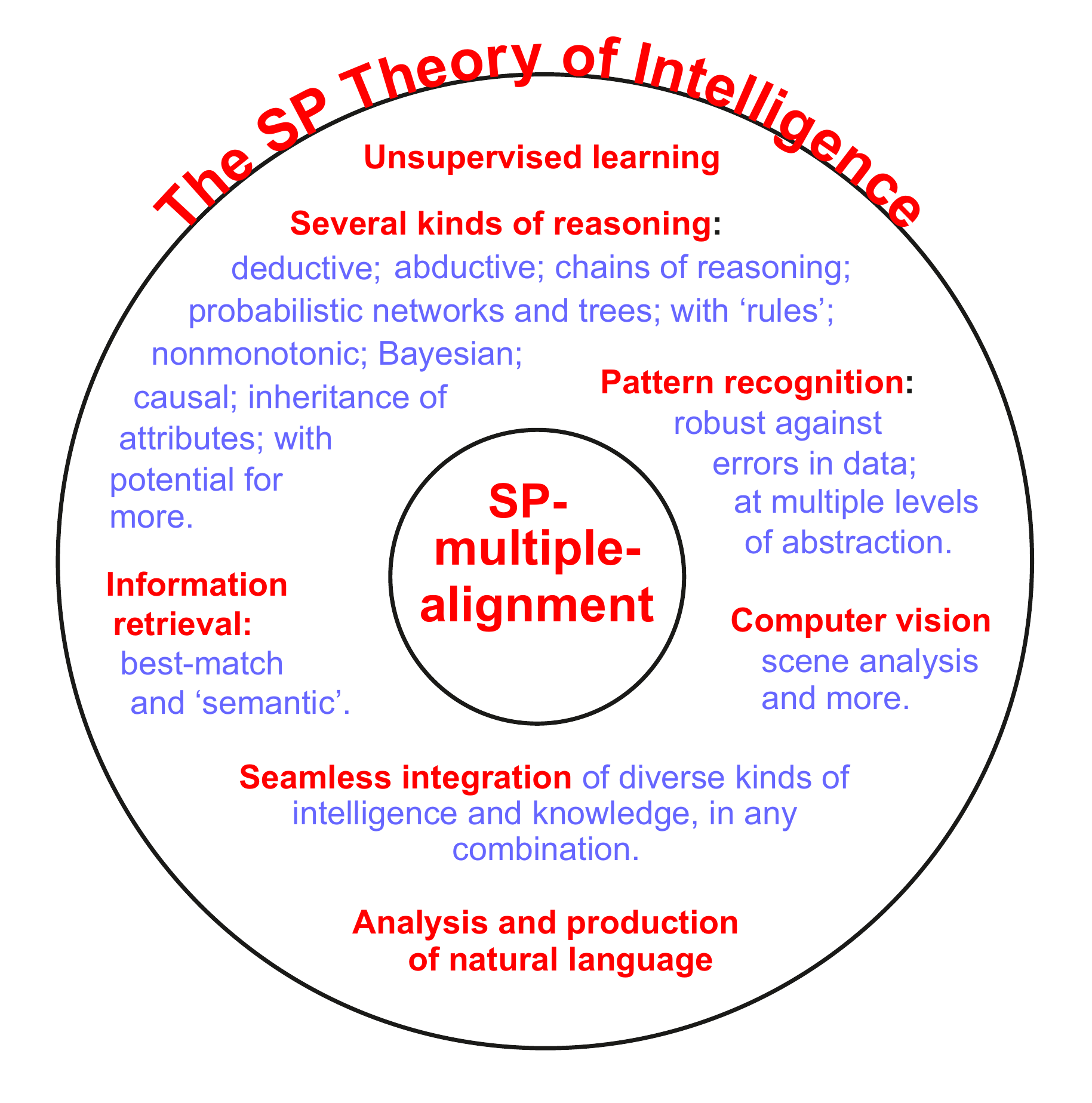}
\caption{A schematic representation of versatility and integration in the SP System, with SPMA centre stage.}
\label{versatility-integration-figure}
\end{figure*}

Figure \ref{versatility-integration-figure} shows schematically how the SP System, with SPMA centre stage, exhibits versatility and integration. The figure is intended to emphasise how development of the SP System has been and is aiming for versatility and integration in the system.

\subsection{Potential benefits and applications of the SP System}\label{benefits-and-applications-appendix}

Apart from its strengths and potential in modelling AI-related functions (Appendix \ref{sp-str-pot-appendix}), it appears that, in more humdrum terms, the SP System has several potential benefits and applications, several of them described in peer-reviewed papers. These include:

\begin{itemize}

    \item {\em Big data}. Somewhat unexpectedly, it has been discovered that the SP System has potential to help solve nine significant problems associated with big data \cite{sp-big-data}. These are: overcoming the problem of variety in big data; the unsupervised learning of structures and relationships in big data; interpretation of big data via pattern recognition, natural language processing; the analysis of streaming data; compression of big data; model-based coding for the efficient transmission of big data; potential gains in computational and energy efficiency in the analysis of big data; managing errors and uncertainties in data; and visualisation of structure in big data and providing an audit trail in the processing of big data.

    \item {\em Autonomous robots}. The SP System opens up a radically new approach to the development of intelligence in autonomous robots \cite{sp-autonomous-robots};

    \item {\em An intelligent database system}. The SP System has potential in the development of an intelligent database system with several advantages compared with traditional database systems \cite{sp-intelligent-database}. In this connection, the SP System has potential to add several kinds of reasoning and other aspects of intelligence to the `database' represented by the World Wide Web, especially if the SP Machine were to be supercharged by replacing the search mechanisms in the foundations of the SP Machine with the high-parallel search mechanisms of any of the leading search engines.

    \item {\em Medical diagnosis}. The SP System may serve as a vehicle for medical knowledge and to assist practitioners in medical diagnosis, with potential for the automatic or semi-automatic learning of new knowledge \cite{sp-medical-diagnosis};

    \item {\em Computer vision and natural vision}. The SP System opens up a new approach to the development of computer vision and its integration with other aspects of intelligence. It also throws light on several aspects of natural vision \cite{sp-vision};

    \item {\em Neuroscience}. As outlined in Appendix \ref{sp-neural-appendix}, abstract concepts in the SP Theory of Intelligence map quite well into concepts expressed in terms of neurons and their interconnections in a version of the theory called {\em SP-Neural} (\cite{spneural-2016}, \cite[Chapter 11]{wolff-2006}). This has potential to illuminate aspects of neuroscience and to suggest new avenues for investigation.

    \item \sloppy {\em Commonsense reasoning}. In addition to the previously-described strengths of the SP System in several kinds of reasoning, the SP System has strengths in the surprisingly challenging area of ``commonsense reasoning'', as described by Ernest Davis and Gary Marcus \cite{davis-marcus-2015}. How the SP System may meet the several challenges in this area is described in \cite{sp-csrk}.

    \item {\em Other areas of application}. The SP System has potential in several other areas of application including \cite{sp-benefits-apps}: the simplification and integration of computing systems; best-match and semantic forms of information retrieval; software engineering \cite{sp-software-engineering}; the representation of knowledge, reasoning, and the semantic web; information compression; bioinformatics; the detection of computer viruses; and data fusion.

    \item {\em Mathematics}. The concept of ICMUP provides an entirely novel interpretation of mathematics \cite{sp-micmup}. This interpretation is quite unlike anything described in existing writings about the philosophy of mathematics or its application in science. There are potential benefits in science from this new interpretation of mathematics.

\end{itemize}

\subsection{Unfinished business and the SP Machine}\label{unfinished-business-appendix}

Like most theories, the SP Theory is not complete. Four pieces of `unfinished business' are described in \cite[Section 3.3]{sp-extended-overview}: 1) The SPCM needs to be generalised to include SP-patterns in two dimensions, with associated processing; 2) Research is needed to discover whether or how the SP concepts may be applied to the identification of low-level perceptual features in speech and images; 3) More work is needed on the development of unsupervised learning in the SPCM; 4) And although the SP Theory has led to the proposal that much of mathematics, perhaps all of it, may be understood as IC \cite{sp-micmup}, research is needed to discover whether or how the SP concepts may be applied in the representation of numbers. A better understanding is also needed of how quantitative concepts such as time, speed, distance, and so on, may be represented in the SP System.

It appears that these problems are soluble and it is anticipated that, with some further research, they can be remedied.

\begin{figure*}[!htbp]
\centering
\includegraphics[width=0.9\textwidth]{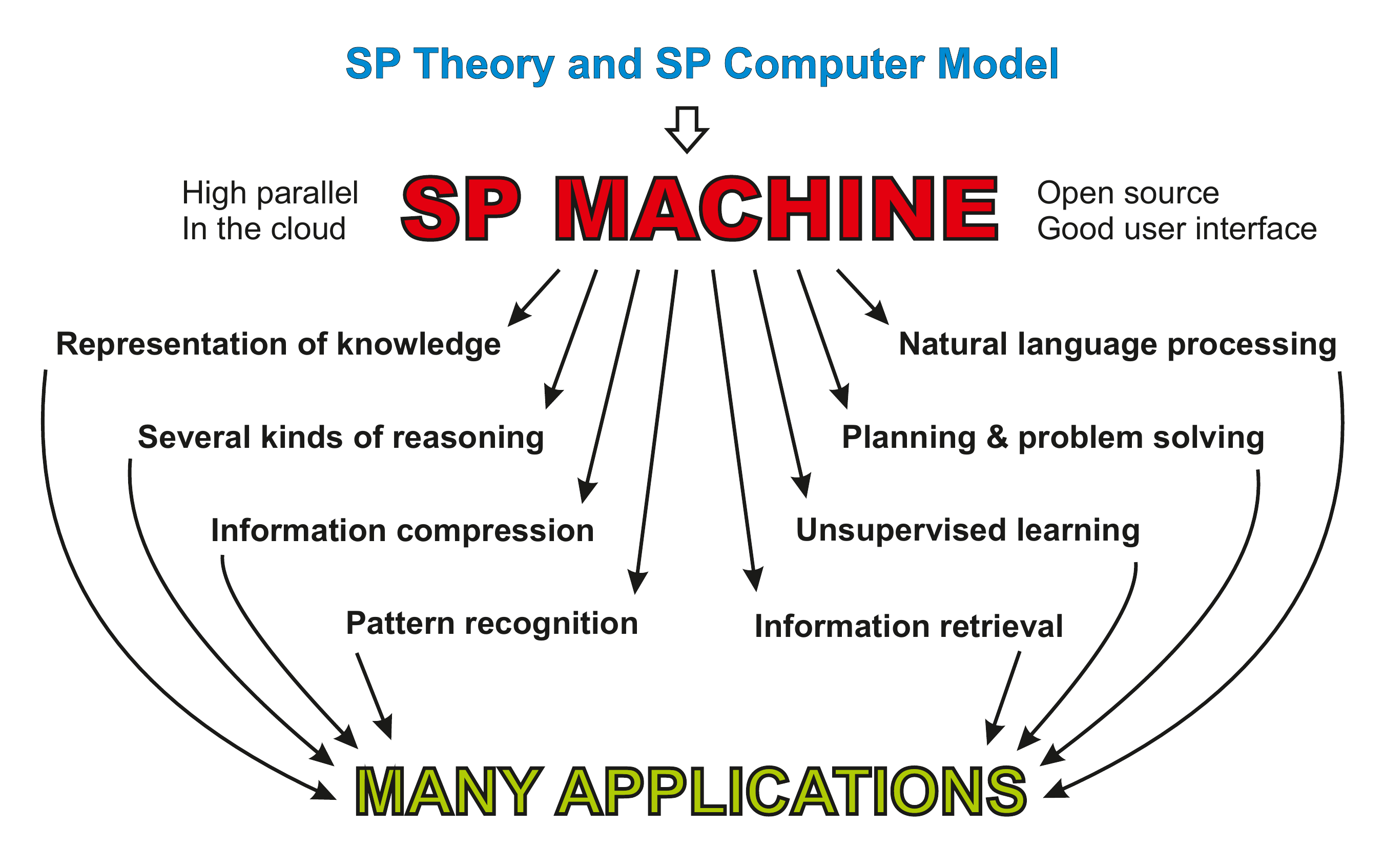}
\caption{Schematic representation of the development and application of the SP Machine. Reproduced from Figure 2 in \cite{sp-extended-overview}, with permission.}
\label{sp-machine-figure}
\end{figure*}

More generally, a programme of research is envisaged, with one or more teams of researchers, or individual researchers, to create a more mature {\em SP Machine}, based on the SPCM, and shown schematically in Figure \ref{sp-machine-figure}. A roadmap for the development of the SP Machine is described in \cite{sp-palade-wolff}.

Elsewhere in this paper, a mature version of the SP Machine may be referred to as an `SPM'.

\end{appendix}

\bibliographystyle{plain}

\end{document}